\documentclass[conf]{new-aiaa} % for conference papers
\usepackage[utf8]{inputenc}
\usepackage{textcomp}

\usepackage{graphicx}
\usepackage{amsmath}
\usepackage[version=4]{mhchem}
\usepackage{siunitx}
\usepackage{longtable,tabularx}
\usepackage{subcaption}

\DeclareMathOperator*{\argmin}{arg\,min}

\usepackage{fancyhdr}
\usepackage{algpseudocode}
\usepackage{algorithm}

\usepackage{hyperref}
\hypersetup{
    colorlinks=true,
    linkcolor=blue,
    filecolor=magenta,      
    urlcolor=cyan,
    pdftitle={Block-structured Operator Inference for coupled multiphysics model reduction},
    }

\title{Block-structured Operator Inference for coupled multiphysics model reduction}

\author{
    Benjamin G. Zastrow \footnote{
        Graduate Research Assistant,
        Department of Aerospace Engineering and Engineering Mechanics,
        bzastrow@utexas.edu,
        AIAA Student Member
        (Corresponding Author).
        },
    Anirban Chaudhuri \footnote{
        Research Scientist,
        Oden Institute for Computational Engineering and Sciences,
        anirbanc@oden.utexas.edu,
        AIAA Senior Member.
        },
    and Karen E. Willcox \footnote{
        Professor, Department of Aerospace Engineering and Engineering Mechanics,
        kwillcox@oden.utexas.edu,
        AIAA Fellow
        (Corresponding Author).
        }
} \affil{University of Texas at Austin, Austin, TX, 78712}
\author{
    Anthony Ashley \footnote{
        Aeronautical Engineer Staff,
        CFD,
        anthony.s.ashley@lmco.com,
        AIAA Member.
        }
    and Michael Chamberlain Henson \footnote{
        Fellow,
        Transformational Solutions/AeroIT,
        mike.c.henson@lmco.com,
        AIAA Associate Fellow.
        }
} \affil{Lockheed Martin, Fort Worth, TX, 76101}

\begin{document}

\footnotetext{Presented as Paper 2023-0330 at the 2023 AIAA SciTech Forum, National Harbor, MD, January 23-27, 2023.}

\maketitle
\thispagestyle{fancy}

\begin{abstract}
This paper presents a block-structured formulation of Operator Inference as a way to learn structured reduced-order models for multiphysics systems.
The approach specifies the governing equation structure for each physics component and the structure of the coupling terms.
Once the multiphysics structure is specified, the reduced-order model is learned from snapshot data following the nonintrusive Operator Inference methodology.
In addition to preserving physical system structure, which in turn permits preservation of system properties such as stability and second-order structure, the block-structured approach has the advantages of reducing the overall dimensionality of the learning problem and admitting tailored regularization for each physics component. 
The numerical advantages of the block-structured formulation over a monolithic Operator Inference formulation are demonstrated for aeroelastic analysis, which couples aerodynamic and structural models.
For the benchmark test case of the AGARD 445.6 wing, block-structured Operator Inference provides an average 20\% online prediction speedup over monolithic Operator Inference across subsonic and supersonic flow conditions in both the stable and fluttering parameter regimes while preserving the accuracy achieved with monolithic Operator Inference.
\end{abstract}

% %%%%%%%%%%%%%%%%%%%%
\section{Introduction}
\label{sec:introduction}
% %%%%%%%%%%%%%%%%%%%%

Building reduced-order models (ROMs) for coupled multiphysics systems requires addressing additional challenges beyond those posed by a single-physics model reduction problem. For example, we often wish to preserve certain features and properties of a single physics regime's physical or numerical model, but model reduction methods that view the system monolithically tend to ignore this type of information.
In this paper, we show how block-structured Operator Inference exploits knowledge of the block structure of coupled multiphysics systems to improve the ROM while preserving the advantages of standard, monolithic Operator Inference.

Model reduction aims to represent high-dimensional dynamics with a lower-dimensional system while maintaining a sufficient level of predictive accuracy. The ROMs can be run at orders of magnitude lower computational cost compared to the full-order models.
Projection-based approaches like the proper orthogonal decomposition (POD) \cite{sirovich1987turbulence, berkooz1993pod, lumley1967structures, rathinam2003newlookpod} use the trajectory data from a full-order model to derive a reduced basis, then the full-order operators of the governing equations are projected onto the reduced basis and the dynamics are integrated forward in time \cite{benner2015pmorsurvey, antoulas2005approximation}.
Another technique, dynamic mode decomposition (DMD) \cite{schmid2010dynamic, Tu2014dynamic}, uses the full-order trajectory data to construct a linear reduced operator to approximate the system, thus enabling analysis of eigenmodes and eigenvalues that often can reasonably characterize even nonlinear full-order dynamics.
However, in the case of multiphysics systems such as aeroelasticity, these types of model reduction methods are often too general to fully embed the knowledge of the nature of the coupling between systems into the ROM, and thus gains in accuracy, robustness, and speed can be missed.

Coupled model reduction methods directly incorporate knowledge of the multiphysics setting into the ROM construction process.
For example, it is common practice in structural dynamics to use component mode synthesis via the Hurty/Craig-Bampton method to break up a system of interacting structural components into separate submodels that only interact along specific sets of boundary degrees of freedom \cite{hurty1960component, craig1968substructures}.
A review of model reduction methods for coupled systems that focuses on linear, time-invariant systems represented via transfer functions is given in \cite{reis2008survey}, focusing on application of balanced truncation and moment matching methods.
Another review of coupled model reduction methods in \cite{benner2015coupled} separates different approaches based on whether the dynamics are coupled through internal states or through inputs and outputs.
In the nonintrusive setting, the authors in \cite{discacciati2024boundary} learn boundary response maps from data to couple models without requiring direct access to the governing equations.
Another nonintrusive approach exploits knowledge of the grid adjacency of a fluid-structure interaction problem to efficiently infer a sparse full-order model, after which POD is used to project the operators to a reduced subspace to construct a ROM \cite{gkimisis2023adjacency}.
From the block-structured perspective, the inherent structure of the governing equations that arises due to coupling is identified in the full-order model structure and then deliberately preserved during construction of the ROM.

Methods for structure-preserving model reduction and coupled model reduction often overlap due to their many shared characteristics. For example, it is common for the governing ODEs of a coupled ROM to possess a particular block structure that can be preserved through the design of a targeted model reduction approach \cite{lall2003structure,beattie2009interpolatory,gugercin2012structure,benner2015coupled, reis2008survey}.
Block structure preservation can be seen as a type of physics-informed modeling because the block structure directly represents the form of each single-physics governing equation and also represents the form of the coupling between the physics regimes.
In this paper, we use another physics-informed method, Operator Inference, to replace the intrusive Galerkin projection step with a nonintrusive linear least squares problem which still targets a ROM form dictated by the full-order model's physics, thus combining the benefits of data-driven learning and physics-informed modeling~\cite{peherstorfer2016opinf,kramer2024learning}.
Another form of structure preservation for the special case of systems governed by nonlinear Lagrangian dynamics appears in \cite{sharma2024spopinf}, where the authors separate the system into linear components learned via Operator Inference and nonlinear components learned via polynomial-augmented multilayer perceptrons, thus utilizing the power of structure-preserving ROMs to embed knowledge of the nonlinear Lagrangian setting into the model. 
In \cite{benner2020nlopinf}, the block structure of the governing PDEs for a batch chromatography system is preserved in a similar manner to this paper by creating a block-structured projection matrix with each block corresponding to the basis for a single subsystem. The authors then infer the linear operators and intrusively impose the known nonlinear terms.

In this paper, we propose a block-structured, coupled multiphysics Operator Inference ROM obtained via a combination of imposition of known structure and nonintrusive inference of unknown operators.
We formulate this approach in the context of a high-dimensional aeroelastic problem where we do not assume the nonlinear terms to be known, and therefore we infer the nonlinear operators nonintrusively as well, which also requires the inclusion of regularization in the inference subproblems.
Incorporating knowledge of a system's block structure into Operator Inference permits greater flexibility than standard, monolithic Operator Inference and enables coupling of multiple physics regimes into a single, multiphysics ROM, while preserving the benefits of Operator Inference's nonintrusive approach. We leverage the added modularity of the block-structured operators to specify the governing equation structure of each physics regime separately, which reduces computational expense by reducing the number of operator terms to be learned.

We demonstrate our approach using the Advisory Group for Aerospace Research and Development (AGARD) 445.6 Wing \cite{yates1987agard} as a high-dimensional coupled multiphysics example application in the aeroelastic setting.
The AGARD wing has been studied frequently in the model reduction context in the literature. In \cite{silva2008simultaneous}, the author uses Walsh functions to simultaneously excite multiple impulse responses in a computational fluid dynamics (CFD) model.
The eigensystem realization algorithm (ERA) \cite{juang1985eigensystem} and nonlinear aeroelastic system identification via Volterra theory \cite{silva2005volterra} are then used to convert these impulses into an unsteady aerodynamic ROM that can be coupled to a structural model to create a coupled aeroelastic ROM \cite{silva2004development, silva2018aerom}.
This Walsh function technique also is used in the POD methodology in \cite{lowe2021efficient}, where the authors build the reduced basis via an incremental approach to avoid handling the full snapshot matrix all at once.
In this work, we take a somewhat different approach and build nonintrusive ROMs for coupled multi-physics problems like the AGARD 445.6 wing via our block-structured Operator Inference method.
Reduced-order modeling has long been identified as an effective way to reduce the cost of aeroelastic analyses and flutter boundary calculations.
Past seminal work has shown the effectiveness of POD-based ROMs for fluid-structure interaction \cite{hall2000pod, dowell2001modeling} and aeroelastic assessments of full aircraft configurations \cite{lieu2006reduced, lieu2007adaptation, amsallem2008interpolation}, vortex lattice models with structural modal decompositions for flutter prediction \cite{hall1994eigenanalysis}, Kriging partial least squares models for aerodynamic design \cite{bouhlel2016kpls}, and the frequency-domain Karhunen-Loeve method for fluid dynamics ROMs \cite{kim1998frequency}.
In this paper, our target is to achieve similar computational cost reductions for aeroelastic analysis, but to do so with nonintrusive methodology, meaning that our approach can be easily applied to legacy and commercial high-fidelity solvers where the user may not have access to the code at the level required for intrusive model reduction approaches.

We generate high-dimensional simulation data using NASA's FUN3D software~\cite{biedron2019fun3d}, where the full-order model consists of a fluid dynamics finite volume model coupled with a structural dynamics finite element modal decomposition model.
FUN3D's aeroelasticity capability couples the structural and fluid dynamics and generates high-fidelity snapshot training data for the AGARD wing.
The main contributions of this paper are:
\begin{enumerate}[label=(\roman*)]
    \item \textit{A formulation and algorithm for block-structured Operator Inference with regularization:}  We exploit the block structure of the governing equations to specify distinct structure and separately tailor the regularization for each regime of the dynamical system.
    \item \textit{Computational speedups due to block-structuring.} Block-structured Operator Inference provides  reduced computational complexity during both inference and prediction. This reduced complexity leads to lower computational costs that improve the efficiency of Operator Inference ROMs.
    \item \textit{Application of block-structured Operator Inference to multiphysics problems.} Block-structured Operator Inference adapts the Operator Inference method to the multiphysics setting by embedding knowledge of the coupling between physics regimes into the learning problem. 
\end{enumerate}

The remainder of this paper is organized as follows. Section~\ref{sec:block-structured-opinf} describes the Operator Inference method, first as initially presented in \cite{peherstorfer2016opinf} and then with the block-structured formulation that is the main contribution of this paper. Section~\ref{sec:agard} presents the AGARD 445.6 wing aeroelastic modeling problem. Section~\ref{sec:analysis} compares the performance of the block-structured and monolithic Operator Inference methods in the context of the AGARD wing. Section~\ref{sec:conclusion} provides concluding remarks.

% %%%%%%%%%%%%%%%%%%%%
\section{Block-structured Operator Inference for multiphysics ROMs}
\label{sec:block-structured-opinf}
% %%%%%%%%%%%%%%%%%%%%

This section develops the block-structured Operator Inference method which is the main methodological contribution of this paper. Sec.~\ref{ssec:system_structure} establishes the aeroelastic modeling problem as a representative multiphysics system. Sec.~\ref{ssec:monolithic_opinf} reviews the standard, monolithic Operator Inference method. Finally, Sec.~\ref{ssec:block_opinf} presents the block-structured Operator Inference approach.

\subsection{Coupled multiphysics system structure}
\label{ssec:system_structure}

We present the aeroelastic modeling problem as a representative example of a block-structured multiphysics system. The block-structured Operator Inference method is applicable to any multiphysics system that contains coupling in an analogous block-structured sense, but the aeroelastic setting will be used throughout the paper for clarity of notation.
Aeroelastic models couple the governing equations of structural dynamics and fluid dynamics together into a single multiphysics dynamical system. The evolution of the state of this system over time is dependent both on the single-physics governing equations and on the manner in which the single-physics governing equations are coupled together. In this paper, the structural dynamics system, fluid dynamics system, and the form of the coupling between the two systems are all permitted to be at most polynomially nonlinear. 
For cases with non-polynomial nonlinear terms, one could employ lifting transformations to introduce auxiliary variables that transform the system to polynomial form (see~\cite{kramer2019nonlinear,qian2020lift}).

\subsubsection{Structural dynamics}
\label{sssec:structural-dynamics}

Structural dynamics are often modeled via linear methods, most notably an eigenvalue decomposition of the governing semi-discrete differential equations (typically derived via a finite element spatial discretization). However, to keep our initial notation general for other systems and methods, we also allow for a constant term and polynomial (here, quadratic) nonlinearities in our structural dynamics governing equations. This leads us to write the general semi-discrete system governing the structural dynamics as
\begin{equation}
    \label{eqn:fom_structural_dynamics_gov_eqns}
    \dot{\mathbf{q}}_\text{s} = \mathbf{c}_\text{s} + \mathbf{A}_\text{s} \mathbf{q}_\text{s} + \mathbf{H}_\text{s} (\mathbf{q}_\text{s} \otimes \mathbf{q}_\text{s}) + \mathbf{F}_\text{s},
\end{equation}
where $\mathbf{q}_\text{s}(t) \in \mathbb{R}^{n_\text{s}}$ is the structural state and $n_\text{s}$ is the dimension of the structural semi-discrete first-order system of ODEs. The operators $\mathbf{c}_\text{s} \in \mathbb{R}^{n_\text{s}}$, $\mathbf{A}_\text{s} \in \mathbb{R}^{n_\text{s} \times n_\text{s}}$, and $\mathbf{H}_\text{s} \in \mathbb{R}^{n_\text{s} \times n_\text{s}^2}$ are the constant, linear, and quadratic structural dynamics operators, respectively, $\mathbf{F}_\text{s}(t) \in \mathbb{R}^{n_\text{s}}$ is the external forcing term, and $\otimes$ denotes the Kronecker product. Note that we employ the compact version of the Kronecker product in the implementation, to avoid computing redundant terms.

\subsubsection{Fluid dynamics}

We similarly employ a quadratic form for the fluid dynamics semi-discrete governing equations:
\begin{equation}
    \label{eqn:fom_fluid_dynamics_gov_eqns}
    \dot{\mathbf{q}}_\text{f} = \mathbf{c}_\text{f} + \mathbf{A}_\text{f} \mathbf{q}_\text{f} + \mathbf{H}_\text{f} (\mathbf{q}_\text{f} \otimes \mathbf{q}_\text{f}) + \mathbf{F}_\text{f},
\end{equation}
where $\mathbf{q}_\text{f}(t) \in \mathbb{R}^{n_\text{f}}$ is the fluid state and $n_\text{f}$ is the dimension of the fluid dynamics semi-discrete system of ODEs. The operators $\mathbf{c}_\text{f} \in \mathbb{R}^{n_\text{f}}$, $\mathbf{A}_\text{f} \in \mathbb{R}^{n_\text{f} \times n_\text{f}}$, and $\mathbf{H}_\text{f} \in \mathbb{R}^{n_\text{f} \times n_\text{f}^2}$ are the constant, linear, and quadratic fluid dynamics operators, respectively, and $\mathbf{F}_\text{f}(t) \in \mathbb{R}^{n_\text{f}}$ is the external forcing term.
As shown in \cite{qian2020lift}, a variety of fluid dynamics models can be written in this quadratic form, particularly with an appropriate choice of fluid state variables. Even for the nonlinear Navier-Stokes equations, there are many situations in which a spatial discretization that yields primarily quadratic nonlinearities can be achieved (one such strategy will be employed in the practical application in Sec.~\ref{sec:agard}). 

\subsubsection{Aeroelastic dynamics}

In keeping with the form established for \eqref{eqn:fom_structural_dynamics_gov_eqns} and \eqref{eqn:fom_fluid_dynamics_gov_eqns}, we express the coupling via the external forcing terms, $\mathbf{F}_\text{s}(t)$ and $\mathbf{F}_\text{f}(t)$. We can assume various forms for these terms, and thus model linear, bilinear, and quadratic coupling. We could also model higher-order polynomial coupling, but for clarity of presentation here we stop at second-order coupling. Thus we write the explicit form of the coupling-specific forcing terms as
\begin{align}
    \label{eqn:fom_structural_coupling_forcing_terms}
    \mathbf{F}_\text{s} &= \mathbf{E}_\text{s} \mathbf{q}_\text{f}
                    + \mathbf{L}_\text{s} (\mathbf{q}_\text{s} \otimes \mathbf{q}_\text{f})
                    + \mathbf{G}_\text{s} (\mathbf{q}_\text{f} \otimes \mathbf{q}_\text{f}) \\
    \label{eqn:fom_fluid_coupling_forcing_terms}
    \mathbf{F}_\text{f} &= \mathbf{E}_\text{f} \mathbf{q}_\text{s}
                    + \mathbf{L}_\text{f} (\mathbf{q}_\text{s} \otimes \mathbf{q}_\text{f})
                    + \mathbf{G}_\text{f} (\mathbf{q}_\text{s} \otimes \mathbf{q}_\text{s}),
\end{align}
where $\mathbf{E}_\text{s}$ and $\mathbf{E}_\text{f}$ are the linear coupling operators, $\mathbf{L}_\text{s}$ and $\mathbf{L}_\text{f}$ are the bilinear coupling operators, and $\mathbf{G}_\text{s}$ and $\mathbf{G}_\text{f}$ are the quadratic coupling operators. Each of these operators acts on quantities that are computed using the state from the other physics component, thus providing the coupling.

Substituting the definitions in \eqref{eqn:fom_structural_coupling_forcing_terms} and \eqref{eqn:fom_fluid_coupling_forcing_terms} into the single-physics governing equations \eqref{eqn:fom_structural_dynamics_gov_eqns} and \eqref{eqn:fom_fluid_dynamics_gov_eqns}, we can combine the equations into a single block-structured multiphysics system of ODEs. This yields the full-order system of multiphysics (aeroelastic) semi-discrete governing equations,
\begin{equation}
    \label{eqn:fom_coupled_gov_eqns}
    \begin{bmatrix}\dot{\mathbf{q}}_\text{s} \\ \dot{\mathbf{q}}_\text{f}\end{bmatrix}
    =
    \begin{bmatrix}\mathbf{c}_\text{s} \\ \mathbf{c}_\text{f}\end{bmatrix}
    +
    \begin{bmatrix}\mathbf{A}_\text{s} & \mathbf{E}_\text{s} \\ \mathbf{E}_\text{f} & \mathbf{A}_\text{f}\end{bmatrix}
    \begin{bmatrix}\mathbf{q}_\text{s} \\ \mathbf{q}_\text{f}\end{bmatrix}
    +
    \begin{bmatrix}\mathbf{H}_\text{s} & \mathbf{L}_\text{s} & \mathbf{G}_\text{s} \\ \mathbf{G}_\text{f} & \mathbf{L}_\text{f} & \mathbf{H}_\text{f}\end{bmatrix}
    \begin{bmatrix}\left(\mathbf{q}_\text{s} \otimes \mathbf{q}_\text{s}\right) \\ \left(\mathbf{q}_\text{s} \otimes \mathbf{q}_\text{f}\right) \\ \left(\mathbf{q}_\text{f} \otimes \mathbf{q}_\text{f}\right)\end{bmatrix}.
\end{equation}
For use in later comparisons to standard Operator Inference, note that \eqref{eqn:fom_coupled_gov_eqns} can be written compactly in monolithic form as
\begin{equation}
    \label{eqn:fom_coupled_gov_eqns_compact}
    \dot{\mathbf{q}} = \mathbf{c} + \mathbf{A} \mathbf{q} + \mathbf{H} (\mathbf{q} \otimes \mathbf{q}),
\end{equation}
where $\mathbf{q} = \begin{bmatrix}\mathbf{q}_\text{s}^\top & \mathbf{q}_\text{f}^\top\end{bmatrix}^\top \in \mathbb{R}^n$ is the coupled state vector, $n=n_\text{s}+n_\text{f}$ is the full order state dimension, and $\mathbf{c} \in \mathbb{R}^n$, $\mathbf{A} \in \mathbb{R}^{n \times n}$, and $\mathbf{H} \in \mathbb{R}^{n \times n^2}$ are respectively the constant, linear, and quadratic monolithic operators.

\subsection{Monolithic Operator Inference}
\label{ssec:monolithic_opinf}

Operator Inference is a nonintrusive, physics-informed reduced-order modeling method that takes advantage of the structure-preserving properties of projection-based model reduction. However, instead of intrusively projecting the semi-discrete operators to the reduced subspace, Operator Inference formulates and solves a linear least squares problem to nonintrusively infer the reduced operators from training data consisting of snapshots of the full-order system state as it is evolved in time \cite{peherstorfer2016opinf}. Here we present an overview of the standard (monolithic) Operator Inference method, applicable to monolithic (single-physics) model-reduction applications (i.e., to systems governed by dynamics of the form of \eqref{eqn:fom_coupled_gov_eqns_compact} that ignore any internal block structure).

The snapshot matrix is assembled by concatenating $k_\text{train}$ full-order training snapshot vectors into a matrix $\mathbf{Q} = [\mathbf{q}_1 \ \mathbf{q}_2 \ \cdots \ \mathbf{q}_{k_\text{train}}] \in \mathbb{R}^{n \times k_\text{train}}$, where each snapshot $\mathbf{q}_i$ with $i = 1, 2, \dots, k_\text{train}$ corresponds to a single column and $k_\text{train}$ refers to the number of training snapshots.
To reduce the dimensionality of the problem, we use the proper orthogonal decomposition (POD) of the full-order snapshot matrix $\mathbf{Q}$ to identify a basis which spans the desired reduced subspace.
POD proceeds by computing the thin singular value decomposition as $\mathbf{Q} = \mathbf{V} \mathbf{\Sigma} \mathbf{W}^\top$, where $\mathbf{V} \in \mathbb{R}^{n \times k_\text{train}}$, $\mathbf{\Sigma} \in \mathbb{R}^{k_\text{train} \times k_\text{train}}$, and $\mathbf{W} \in \mathbb{R}^{k_\text{train} \times k_\text{train}}$. We then select the first $r$ left singular vectors $\mathbf{v}_j$ with $j = 1, 2, \dots, r$ which correspond to the first $r$ columns of $\mathbf{V}$, to give the basis matrix $\mathbf{V}_r= [\mathbf{v}_1 \ \mathbf{v}_2 \ \cdots \ \mathbf{v}_r] \in \mathbb{R}^{n \times r}$, where $r$ is the desired reduced state dimension.

The full-order state is approximated in the reduced basis as $\mathbf{q} \approx \mathbf{V}_r \widehat{\mathbf{q}}$, where $\widehat{\mathbf{q}} \in \mathbb{R}^r$ is the reduced state vector. 
We now seek a ROM that governs the dynamics of the reduced state  $\widehat{\mathbf{q}}(t)$.
Projection-based model reduction theory defines the form of the ROM to be
\begin{equation}
    \label{eqn:rom_mono}
    \dot{\widehat{\mathbf{q}}} = \widehat{\mathbf{c}} + \widehat{\mathbf{A}} \widehat{\mathbf{q}} + \widehat{\mathbf{H}} \left(\widehat{\mathbf{q}} \otimes \widehat{\mathbf{q}}\right)
\end{equation}
where $\widehat{\mathbf{c}} \in \mathbb{R}^r$, $\widehat{\mathbf{A}} \in \mathbb{R}^{r \times r}$, and $\widehat{\mathbf{H}} \in \mathbb{R}^{r \times r^2}$ are the constant, linear, and quadratic monolithic reduced operators. Note that the ROM \eqref{eqn:rom_mono} has the same structure as the full-order model \eqref{eqn:fom_coupled_gov_eqns_compact}, due to the properties of approximation by projection (see \cite{peherstorfer2016opinf,kramer2024learning}).

We project the snapshot matrix to the reduced space defined by the POD basis, $\widehat{\mathbf{Q}} = \mathbf{V}_r^\top \mathbf{Q} \in \mathbb{R}^{r \times k_\text{train}}$.
We then nonintrusively infer the reduced operators by solving the linear least squares problem:
\begin{equation}
    \label{eqn:lstsq}
    \widehat{\mathbf{O}} = \argmin_{\widehat{\mathbf{O}}} \left\|\widehat{\mathbf{D}} \widehat{\mathbf{O}}^\top - \widehat{\mathbf{R}}^\top\right\|_F^2
\end{equation}
where $F$ denotes the Frobenius norm and we define
\begin{align*}
    \widehat{\mathbf{D}} &= \begin{bmatrix}\mathbf{1}_{k_\text{train}}^\top & \widehat{\mathbf{Q}}^\top & \left(\widehat{\mathbf{Q}} \otimes \widehat{\mathbf{Q}}\right)^\top\end{bmatrix} &\quad\text{data matrix}\quad \\
    \widehat{\mathbf{O}} &= \begin{bmatrix}\widehat{\mathbf{c}} & \widehat{\mathbf{A}} & \widehat{\mathbf{H}}\end{bmatrix} &\quad\text{operator matrix}\quad \\
    \widehat{\mathbf{R}} &= \begin{bmatrix}\dot{\widehat{\mathbf{Q}}}\end{bmatrix} &\quad\text{right-hand-side matrix}\quad
\end{align*}
where $\mathbf{1}_{k_\text{train}} \in \mathbb{R}^{1 \times k_\text{train}}$ is a $k_\text{train}$-dimensional unity row vector.
If the time derivatives from the solver are available, then we populate the right-hand-side matrix, $\widehat{\mathbf{R}}$, directly from the reduced time derivative snapshot matrix provided by the flow solver, $\dot{\widehat{\mathbf{Q}}} = \mathbf{V}_r^\top \dot{\mathbf{Q}}$. However, these values are often not available when working with commercial or restricted solvers (such as FUN3D, which we use for the numerical example beginning in Sec. \ref{sec:agard}). Thus, our recourse in this work is to compute $\dot{\widehat{\mathbf{Q}}}$ via a sixth-order centered finite difference approximation. Note that we compute the finite differences in the reduced subspace, not the full space, to avoid the additional computational cost of a second projection operation.

For many real-world Operator Inference applications with high-dimensional full-order models and complex dynamics, the stability of the inferred ROMs can be a challenge.
This is due to the use of sparse and noisy training data, which, when compounded with ill-conditioned least squares problems, can lead to overfitting of the operators. Therefore, adding a regularizer $\mathcal{R}$ to \eqref{eqn:lstsq} is typically needed to improve the robustness of the learning step at the expense of the possible loss of some accuracy \cite{mcquarrie2021data}.
Using  a Tikhonov regularization function $\mathcal{R}(\widehat{\mathbf{O}}) = \gamma_c \left\|\widehat{\mathbf{c}}\right\|_F^2 + \gamma_A \left\|\widehat{\mathbf{A}}\right\|_F^2 + \gamma_H \left\|\widehat{\mathbf{H}}\right\|_F^2$, the least squares problem then becomes
\begin{equation}
    \label{eqn:lstsq_reg}
    \widehat{\mathbf{O}}
    = \argmin_{\widehat{\mathbf{O}}}
    \left(
    \left\| \widehat{\mathbf{D}} \widehat{\mathbf{O}}^\top - \widehat{\mathbf{R}}^\top \right\|_F^2
    + \gamma_c \left\| \widehat{\mathbf{c}} \right\|_F^2
    + \gamma_A \left\| \widehat{\mathbf{A}} \right\|_F^2
    + \gamma_H \left\| \widehat{\mathbf{H}} \right\|_F^2
    \right),
\end{equation}
where $\gamma_c$, $\gamma_A$, and $\gamma_H$ are scalar regularization hyperparameters that we choose to separately weight the penalization of each operator. In practice, for Operator Inference these regularization hyperparameters are identified by performing a grid search to identify the combination of regularization levels that leads to ROM operators which most accurately predict the true training snapshot trajectory \cite{mcquarrie2021data}.
One may also incorporate a bounded growth constraint \cite{mcquarrie2021data} to further encourage the stability of the resulting ROMs as described in Appendix A of \cite{qian2022pdes}. This constraint requires that the deviations of each reduced state do not exceed the training regime deviations by more than a user-determined factor, $\alpha$, which becomes another hyperparameter for the learning problem that can be identified via a similar grid search.

\subsection{Block-structured Operator Inference}
\label{ssec:block_opinf}

Block-structured Operator Inference uses multiple instances of the monolithic Operator Inference approach presented in Sec.~\ref{ssec:monolithic_opinf} to decompose a block-structured model reduction problem into multiple smaller subproblems. These subproblems are related to each other via the coupling mechanisms discussed in Sec.~\ref{ssec:system_structure}. The single-physics operators corresponding to each physics regime (e.g., structural dynamics and fluid dynamics) are inferred via separate least squares subproblems. This separation allows for added flexibility in regularizing the operators, nonintrusively imposing structure, and intrusively imposing any known operators for each physics regime.

To define the decomposition of the system into subproblems, consider the multiphysics governing equations with block structure given by \eqref{eqn:fom_coupled_gov_eqns}. Separate bases $\mathbf{V}_\text{s} \in \mathbb{R}^{r_\text{s} \times n_\text{s}}$ and $\mathbf{V}_\text{f} \in \mathbb{R}^{r_\text{f} \times n_\text{f}}$ must be chosen for each subproblem (e.g., via POD), where $r_\text{s}$ and $r_\text{f}$ are respectively the dimensions of the reduced systems of ODEs for the structural dynamics and fluid dynamics. These bases are combined in block-diagonal form into a coupled basis
\begin{equation}
    \label{eqn:coupled_basis}
    \mathbf{V}_r = \begin{bmatrix}\mathbf{V}_\text{s} & \\ & \mathbf{V}_\text{f}\end{bmatrix}
\end{equation}
in order to maintain the desired structure-preserving behavior \cite{schilders2014}.
This form of the coupled basis leads to an analogous structure-preserving projection property to that discussed in Sec.~\ref{ssec:monolithic_opinf}, which in turn means that the resulting multiphysics ROM structure is
\begin{equation}
    \label{eqn:rom_block}
    \begin{bmatrix}\dot{\widehat{\mathbf{q}}}_\text{s} \\ \dot{\widehat{\mathbf{q}}}_\text{f}\end{bmatrix}
    =
    \begin{bmatrix}\widehat{\mathbf{c}}_\text{s} \\ \widehat{\mathbf{c}}_\text{f}\end{bmatrix}
    +
    \begin{bmatrix}\widehat{\mathbf{A}}_\text{s} & \widehat{\mathbf{E}}_\text{s} \\ \widehat{\mathbf{E}}_\text{f} & \widehat{\mathbf{A}}_\text{f}\end{bmatrix}
    \begin{bmatrix}\widehat{\mathbf{q}}_\text{s} \\ \widehat{\mathbf{q}}_\text{f}\end{bmatrix}
    +
    \begin{bmatrix}\widehat{\mathbf{H}}_\text{s} & \widehat{\mathbf{L}}_\text{s} & \widehat{\mathbf{G}}_\text{s} \\ \widehat{\mathbf{G}}_\text{f} & \widehat{\mathbf{L}}_\text{f} & \widehat{\mathbf{H}}_\text{f}\end{bmatrix}
    \begin{bmatrix}\left(\widehat{\mathbf{q}}_\text{s} \otimes \widehat{\mathbf{q}}_\text{s}\right) \\ \left(\widehat{\mathbf{q}}_\text{s} \otimes \widehat{\mathbf{q}}_\text{f}\right) \\ \left(\widehat{\mathbf{q}}_\text{f} \otimes \widehat{\mathbf{q}}_\text{f}\right)\end{bmatrix},
\end{equation}
which matches the structure of the full-order block-structured governing ODEs in \eqref{eqn:fom_coupled_gov_eqns}.
In other words, our block-structured Operator Inference approach yields a ROM that retains the block structure of the original multiphysics full-order system. 

To determine the ROM operators, we observe that the operators for each physics component can be inferred via a separate least squares process of an analogous form to \eqref{eqn:lstsq_reg}. This is due to the separability of each column of a matrix least squares problem into an independent vector least squares problem (see \cite{peherstorfer2016opinf} for details).
Therefore, for the structural dynamics subproblem, we solve
\begin{equation}
    \label{eqn:lstsq_struct}
    \widehat{\mathbf{O}}_\text{s} = \argmin_{\widehat{\mathbf{O}}_\text{s}} \left\| \widehat{\mathbf{D}}_\text{s} \widehat{\mathbf{O}}_\text{s}^\top - \widehat{\mathbf{R}}_\text{s}^\top\right\|_F^2 + \mathcal{R}(\widehat{\mathbf{O}}_\text{s}),
\end{equation}
where we define
\begin{align*}
    \widehat{\mathbf{D}}_\text{s} &= \begin{bmatrix} \mathbf{1}_k^\top & \widehat{\mathbf{Q}}_\text{s}^\top & \widehat{\mathbf{Q}}_\text{f}^\top & \left(\widehat{\mathbf{Q}}_\text{s} \otimes \widehat{\mathbf{Q}}_\text{s}\right)^\top & \left(\widehat{\mathbf{Q}}_\text{s} \otimes \widehat{\mathbf{Q}}_\text{f}\right)^\top & \left(\widehat{\mathbf{Q}}_\text{f} \otimes \widehat{\mathbf{Q}}_\text{f}\right)^\top \end{bmatrix} &\quad\text{data matrix}\quad \\
    \widehat{\mathbf{O}}_\text{s} &= \begin{bmatrix}\widehat{\mathbf{c}}_\text{s} & \widehat{\mathbf{A}}_\text{s} & \widehat{\mathbf{E}}_\text{s} & \widehat{\mathbf{H}}_\text{s} & \widehat{\mathbf{L}}_\text{s} & \widehat{\mathbf{G}}_\text{s} \end{bmatrix} &\quad\text{operator matrix}\quad \\
    \widehat{\mathbf{R}}_\text{s} &= \begin{bmatrix}\dot{\widehat{\mathbf{Q}}}_\text{s}\end{bmatrix} &\quad\text{right-hand-side matrix}\quad
\end{align*}
For the fluid dynamics subproblem, we have
\begin{equation}
    \label{eqn:lstsq_fluid}
    \widehat{\mathbf{O}}_\text{f} = \argmin_{\widehat{\mathbf{O}}_\text{f}} \left\|\widehat{\mathbf{D}}_\text{f} \widehat{\mathbf{O}}_\text{f}^\top - \widehat{\mathbf{R}}_\text{f}^\top\right\|_F^2 + \mathcal{R}(\widehat{\mathbf{O}}_\text{f}),
\end{equation}
where we define
\begin{align*}
    \widehat{\mathbf{D}}_\text{f} &= \begin{bmatrix} \mathbf{1}_k^\top & \widehat{\mathbf{Q}}_\text{s}^\top & \widehat{\mathbf{Q}}_\text{f}^\top & \left(\widehat{\mathbf{Q}}_\text{s} \otimes \widehat{\mathbf{Q}}_\text{s}\right)^\top & \left(\widehat{\mathbf{Q}}_\text{s} \otimes \widehat{\mathbf{Q}}_\text{f}\right)^\top & \left(\widehat{\mathbf{Q}}_\text{f} \otimes \widehat{\mathbf{Q}}_\text{f}\right)^\top \end{bmatrix} &\quad\text{data matrix}\quad \\
    \widehat{\mathbf{O}}_\text{f} &= \begin{bmatrix}\widehat{\mathbf{c}}_\text{f} & \widehat{\mathbf{E}}_\text{f} & \widehat{\mathbf{A}}_\text{f} & \widehat{\mathbf{G}}_\text{f} & \widehat{\mathbf{L}}_\text{f} & \widehat{\mathbf{H}}_\text{f}\end{bmatrix} &\quad\text{operator matrix}\quad \\
    \widehat{\mathbf{R}}_\text{f} &= \begin{bmatrix}\dot{\widehat{\mathbf{Q}}}_\text{f}\end{bmatrix} &\quad\text{right-hand-side matrix}\quad
\end{align*}

Note that in this general formulation, $\widehat{\mathbf{D}}_\text{s} = \widehat{\mathbf{D}}_\text{f}$, and the structural and fluid Operator Inference problems differ only in the right-hand-side matrices
since $\widehat{\mathbf{R}}_\text{s} \neq \widehat{\mathbf{R}}_\text{f}$. 
However, for specific problems, the multiphysics ROM \eqref{eqn:rom_block} may not have all blocks populated, which would lead to $\widehat{\mathbf{D}}_\text{s} \neq \widehat{\mathbf{D}}_\text{f}$. As we will show in  Sec.~\ref{sec:agard}, introducing problem-specific structure into the form of the ROM is part of the utility and desirability of the block-structured approach.

Regularization for each least squares subproblem can be achieved in the same manner as for the monolithic case in Sec.~\ref{ssec:monolithic_opinf}. The scales of the regularization values may differ for each subproblem.
This gives an opportunity to better tune the regularization strategy, but also provides a potential computational challenge, since the increasing dimensionality of the regularization parameters leads to increased computational complexity in determining optimal regularization levels.

% %%%%%%%%%%%%%%%%%%%%
\section{Aeroelastic model of the AGARD 445.6 wing}
\label{sec:agard}
% %%%%%%%%%%%%%%%%%%%%

The AGARD 445.6 wing \cite{yates1987agard} is a commonly used aeroelastic modeling validation case.
Experimental flutter testing was conducted on the AGARD wing at NASA's Transonic Dynamics Tunnel \cite{yates1987agard} and the results of those tests have become a canonical flutter modeling validation dataset.
This section describes the full-order numerical model that will generate train and test data sets to be used to study the Operator Inference ROMs. Section~\ref{ssec:agard_geom_grid} describes the geometry and spatial discretizations of the AGARD 445.6 wing, and Sec.~\ref{ssec:numerical_model} presents the full-order aeroelastic modeling approach as implemented in FUN3D to couple the fluid dynamics and structural dynamics.

\subsection{Geometry and discretization}
\label{ssec:agard_geom_grid}

The AGARD wing is a half-span swept and tapered wing with a root chord of 22 inches, a tip chord of 14.5 inches, a taper ratio of 0.6576, an aspect ratio of 1.6525, a span of 30 inches, and a quarter-chord sweepback angle of $45$ degrees.
It has a symmetric NACA65A004 airfoil and is made of laminated mahogany that was deliberately weakened by drilling holes and filling them with foam, inducing heightened flexibility during wind tunnel testing \cite{yates1987agard}.

The CFD grid used in this paper\footnote{Provided by Pawel Chwalowski of the NASA Langley Research Center.
}
was designed for a viscous, turbulent finite volume CFD model and has $g_f = 3,613,171$ grid points. The structural dynamics grid\footnote{FUN3D v13.4 Training - Session 16: Aeroelastic Simulations: \url{https://fun3d.larc.nasa.gov/session16_2018.pdf}}
uses two-dimensional plate finite elements and has $g_s = 121$ grid points.

\subsection{Numerical model}
\label{ssec:numerical_model}

We use NASA's FUN3D solver as the full-order model. FUN3D is a fully-unstructured Navier-Stokes finite volume solver for CFD \cite{biedron2019fun3d}. It has an integrated aeroelasticity capability that couples the high-fidelity nonlinear fluid dynamics solution to a linear structural dynamics solver based on an eigenvalue modal decomposition \cite{biedron2009recent}.
Note that
we refer to this setup as our full-order model, even though the structural dynamics model is already a reduced-order modal model.
This eigenvalue modal decomposition approach, which will be described in detail in Sec.~\ref{sssec:sd_fom}, uses the eigenvectors of the finite element model to span the structural reduced basis $\mathbf{V}_\text{s}$ from \eqref{eqn:coupled_basis} and thus provide us with a structural dynamics ROM.
This is acceptable in the context of aeroelastic modeling because the fluid dynamics solution typically requires a significant majority of the computational resources and is therefore the component of the solution that is of most interest in a model reduction context. Here we first present the fluid dynamics model (Sec.~\ref{sssec:fd_fom}), then the structural dynamics model (Sec.~\ref{sssec:sd_fom}), and finally the coupling between the two models (Sec.~\ref{sssec:coupled_fom}).

\subsubsection{Fluid dynamics}
\label{sssec:fd_fom}

We use FUN3D to solve the unsteady, compressible, turbulent Navier-Stokes equations for flow over the AGARD 445.6 wing, starting from the form
\begin{subequations}
    \label{eqn:ns_fun3d}
    \begin{equation}
        \frac{\partial \rho}{\partial t} + \boldsymbol{\nabla} \boldsymbol{\cdot} \rho \vec{u} = 0
    \end{equation}
    \begin{equation}
        \frac{\partial \rho \vec{u}}{\partial t} + \boldsymbol{\nabla} \boldsymbol{\cdot} (\rho \vec{u} \otimes \vec{u} - \boldsymbol{\sigma})  = 0
    \end{equation}
    \begin{equation}
        \frac{\partial e}{\partial t} + \boldsymbol{\nabla} \boldsymbol{\cdot} (e \vec{u} - k_\text{thermal} \nabla T - \boldsymbol{\sigma} \cdot \vec{u}) = 0
    \end{equation}
\end{subequations}
where $\boldsymbol{\sigma} = -p \mathbf{I} + \mu \left[-\frac{2}{3} (\boldsymbol{\nabla} \boldsymbol{\cdot} \vec{u}) \mathbf{I} + \nabla \vec{u} + (\nabla \vec{u})^\top\right]$ is the diffusive flux tensor, $p$ is the pressure, $\rho$ is the density, $\vec{u} = [u \ v \ w]$ are the fluid velocity components in each coordinate direction, $e$ is the energy per unit volume, and $k_\text{thermal}$ is the thermal conductivity. One must also specify suitable initial conditions and boundary conditions. 

We assume a thermally and calorically perfect gas with equation of state
\begin{equation}
    T = \frac{p}{\rho R}
\end{equation}
where $T$ is the temperature and $R=1716.49$ ft-lbf/slug-\si{\degree}R is the specific gas constant for air \cite{chen2017foundations}. The speed of sound is defined by the relation
\begin{equation}
    a^2 = \gamma_\text{specific} R T
\end{equation}
where $\gamma_\text{specific}=1.4$ is the specific heat ratio for air. The specific heat relations are
\begin{align}
    \gamma_\text{specific} &= \frac{c_p}{c_v} \\
    c_p - c_v &= R \\
    \frac{R}{c_p} &= \frac{\gamma_\text{specific}-1}{\gamma_\text{specific}}
\end{align}
where $c_p$ and $c_v$ are the specific heat capacities at constant pressure and constant volume. Finally, we define the relationship between temperature and viscosity via Sutherland's law
\begin{equation}
    \mu = \mu_\text{ref} \frac{T_\text{ref}+C}{T+C} \left(\frac{T}{T_\text{ref}}\right)^{3/2}
\end{equation}
where $C=198.6$~\si{\degree}R for air \cite{biedron2019fun3d}, $T_\text{ref}=518.69$~\si{\degree}R (assuming standard atmosphere at sea level), and $\mu_\text{ref}=3.737E-7$ slug/ft-s (assuming standard atmosphere at sea level) \cite{etkin1996dynamics}.

To close the RANS equations for turbulent (viscous) flow, we use the one equation Spalart-Allmaras turbulence model \cite{spalart1994turbulence}. Adding this turbulence model means that $\mu = \mu_l + \mu_t$, where $\mu_l$ is the laminar viscosity computed via Sutherland's law and $\mu_t$ is the turbulent eddy viscosity computed via the solution of the Spalart-Allmaras turbulence equation. In practice, FUN3D solves the nondimensional form of \eqref{eqn:ns_fun3d}. Thus, instead of assuming one of the dimensional parameters listed above, we close the nondimensional governing equations by assuming a Prandtl number of $P_r=0.72$ for air, where $P_r=\frac{c_p \mu}{k_\text{thermal}}$. See \cite{biedron2019fun3d} for further details about the nondimensionalization.

FUN3D uses the finite volume method to spatially discretize \eqref{eqn:ns_fun3d} and outputs the discrete solution in terms of primitive variables: pressure, $\mathbf{p}$, component fluid velocities, $\mathbf{u}$, $\mathbf{v}$, and $\mathbf{w}$, and density $\boldsymbol{\rho}$. We use the $g_f$-dimensional grid mentioned above,
so with five fluid states at each grid point, the full-order CFD model has $n_\text{f} = 5g_f = 18,065,855$ fluid degrees of freedom for the Navier-Stokes equations plus another $g_f = 3,613,171$ degrees of freedom for the turbulence model. For the purposes of Operator Inference, we target the form of the PDE governing equations rather than their discretized RANS form, thus we will neglect the turbulence variable's degrees of freedom in our ROM formulation. See Sec. \ref{ssec:preprocessing} for more details.
The effects of the structural motion of the wing enter through boundary conditions, which in turn introduce the external forcing term $\mathbf{F}_\text{f}(t)$ on the right-hand-side of the semi-discrete fluid dynamics governing equation \eqref{eqn:fom_fluid_dynamics_gov_eqns}.

\subsubsection{Structural dynamics}
\label{sssec:sd_fom}

\begin{figure}[!htb]
    \centering
    \begin{subfigure}{1.625in}
        \centering
        \includegraphics[width=\textwidth]{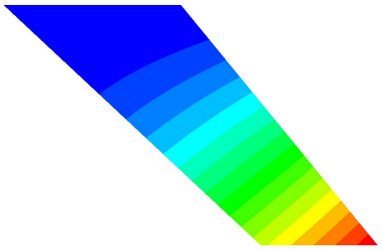}
        \caption{Mode 1 (9.6 Hz)}
    \end{subfigure}
    \begin{subfigure}{1.625in}
        \centering
        \includegraphics[width=\textwidth]{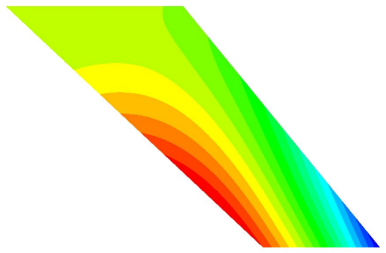}
        \caption{Mode 2 (38.2 Hz)}
    \end{subfigure}
    \\
    \begin{subfigure}{1.625in}
        \centering
        \includegraphics[width=\textwidth]{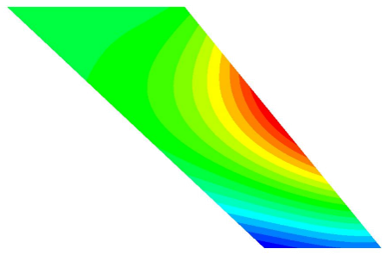}
        \caption{Mode 3 (48.3 Hz)}
    \end{subfigure}
    \begin{subfigure}{1.625in}
        \centering
        \includegraphics[width=\textwidth]{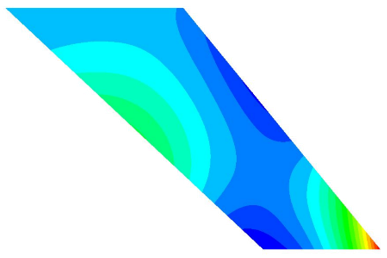}
        \caption{Mode 4 (91.5 Hz)}
    \end{subfigure}
    \caption{Structural mode shapes of the AGARD wing, scaled independently for visualization.}
    \label{fig:agard_modes}
\end{figure}

The finite element method is used to spatially discretize the AGARD wing geometry and generate a second order, linear system of governing equations for the structural dynamics of the form
\begin{equation}
    \mathbf{M} \ddot{\boldsymbol{\delta}} + \mathbf{C} \dot{\boldsymbol{\delta}} + \mathbf{K} \boldsymbol{\delta} = \mathbf{F}_\delta
\end{equation}
where $\boldsymbol{\delta}$ is the state vector of displacements, $\mathbf{M}$, $\mathbf{C}$, and $\mathbf{K}$ are the mass, damping, and stiffness matrices, respectively, and $\mathbf{F}_\delta$ is the external input forcing term. To identify a reduced basis for the structural dynamics, we compute an eigenvalue modal decomposition via the generalized eigenvalue problem
\begin{equation}
    \mathbf{K} \boldsymbol{\phi}_i = \omega^2 \mathbf{M} \boldsymbol{\phi}_i
\end{equation}
where $\boldsymbol{\phi}_i$ is the eigenvector corresponding to the $i^\text{th}$ eigenvalue $\omega_i$. Each eigenvector $\boldsymbol{\phi}_i$ corresponds to a global structural mode shape (reduced degree of freedom), of which the first four are shown in Fig.~\ref{fig:agard_modes} and are taken from
\cite{yates1987agard} (formatted for FUN3D in the FUN3D tutorial materials \footnote{FUN3D v13.4 Training - Session 16: Aeroelastic Simulations: \url{https://fun3d.larc.nasa.gov/session16_2018.pdf}}).

Concatenating the eigenvectors (i.e., the basis vectors) produces a basis matrix $\boldsymbol{\Phi} \in \mathbb{R}^{n_\text{s} \times n_\text{s}}$, which can then be truncated to the desired reduced-order dimension $r_\text{s}$ by keeping only the first $r_\text{s}$ columns. We denote this reduced basis matrix $\mathbf{V}_\text{s} \in \mathbb{R}^{n_\text{s} \times r_\text{s}}$ to indicate that it spans the reduced subspace for the structural dynamics. Now the structural dynamics equation is reduced via Galerkin projection by defining the reduced structural state $\boldsymbol{\eta}$ via the projection $\boldsymbol{\delta} = \mathbf{V}_\text{s} \boldsymbol{\eta}$ and left multiplying by $\mathbf{V}_\text{s}^\top$ to get
\begin{equation}
    \mathbf{V}_\text{s}^\top \mathbf{M} \mathbf{V}_\text{s} \ddot{\boldsymbol{\eta}} + \mathbf{V}_\text{s}^\top \mathbf{C} \mathbf{V}_\text{s} \dot{\boldsymbol{\eta}} + \mathbf{V}_\text{s}^\top \mathbf{K} \mathbf{V}_\text{s} \boldsymbol{\eta} = \mathbf{V}_\text{s}^\top \mathbf{F}_{\boldsymbol{\delta}}.
\end{equation}
Due to the orthogonality properties of the eigenvectors, we have $\mathbf{V}_\text{s}^\top \mathbf{M} \mathbf{V}_\text{s} = \mathbf{I}$, $\mathbf{V}_\text{s}^\top \mathbf{C} \mathbf{V}_\text{s} = \boldsymbol{\zeta}$, $\mathbf{V}_\text{s}^\top \mathbf{K} \mathbf{V}_\text{s} = \boldsymbol{\Omega}$, and $\mathbf{F}_{\boldsymbol{\eta}} = \mathbf{V}_\text{s}^\top \mathbf{F}_{\boldsymbol{\delta}}$, so we simplify to obtain
\begin{equation}
    \ddot{\boldsymbol{\eta}} + \boldsymbol{\zeta} \dot{\boldsymbol{\eta}} + \boldsymbol{\Omega} \boldsymbol{\eta} = \mathbf{F}_{\boldsymbol{\eta}}
\end{equation}
where
$\boldsymbol{\zeta} = \text{diag} \left( 2 \omega_1 \zeta_1, \ 2 \omega_2 \zeta_2, \ \dots, \ 2 \omega_{n_s} \zeta_{n_s} \right)$
and
$\boldsymbol{\Omega} = \text{diag} \left(\omega_1^2, \ \omega_2^2, \ \dots, \ \omega_{n_s}^2 \right)$ are diagonal matrices.
Then we convert the system of uncoupled scalar second-order equations to a system of vector first order equations by redefining the reduced structural state as $\widehat{\mathbf{q}}_\text{s} = [\boldsymbol{\eta}^\top \ \dot{\boldsymbol{\eta}}^\top]^\top$. This produces a first-order, block-structured system of ODEs as desired, written as
\begin{equation}
    \begin{bmatrix}\dot{\boldsymbol{\eta}} \\ \ddot{\boldsymbol{\eta}}\end{bmatrix}
    =
    \begin{bmatrix}\mathbf{0} & \mathbf{I} \\ -\boldsymbol{\Omega} & -\boldsymbol{\zeta}\end{bmatrix}
    \begin{bmatrix}\boldsymbol{\eta} \\ \dot{\boldsymbol{\eta}}\end{bmatrix}
    +
    \begin{bmatrix}\mathbf{0} \\ \mathbf{F}_{\boldsymbol{\eta}}\end{bmatrix}
\end{equation}
where for this work, we assume zero structural damping for all of the modes, $\zeta_i = 0.0$. This system of ODEs can be written compactly as
\begin{equation}
    \label{eqn:fom_ode_struct}
    \dot{\widehat{\mathbf{q}}}_\text{s} = \widehat{\mathbf{A}}_\text{s} \widehat{\mathbf{q}}_\text{s} + \widehat{\mathbf{F}}_\text{s}.
\end{equation}

\subsubsection{Aeroelastic coupling}
\label{sssec:coupled_fom}

FUN3D's aeroelasticity capability couples the modal structural dynamics model and the RANS CFD model to create an aeroelastic multiphysics simulation \cite{biedron2009recent}. The inputs from the fluid dynamics to the structural dynamics enter via the generalized aerodynamic forces $\mathbf{F}_\text{s}$ from a given timestep of the CFD solution. Conversely, the inputs from the structural dynamics to the fluid dynamics enter via specification of fluid velocities at the fluid-solid interfaces, along with a linear elasticity mesh deformation approach which modifies the spatial discretization before each new timestep of the CFD solution.
We refer the reader to \cite{biedron2019fun3d} and \cite{biedron2009recent} for further details regarding the FUN3D and aeroelasticity implementations.

% %%%%%%%%%%%%%%%%%%%%
\section{Block-structured Operator Inference for the AGARD 445.6 wing}
\label{sec:analysis}
% %%%%%%%%%%%%%%%%%%%%

Section~\ref{ssec:data_generation} summarizes the flow conditions used to generate full-order model training data for the Operator Inference ROM.
Section~\ref{ssec:preprocessing} describes the preprocessing steps performed to prepare the fluid training data for Operator Inference, in particular the state variable transformation, shifting, and scaling of the snapshots.
Section~\ref{ssec:svd} studies the singular value decomposition of the fluid dynamics snapshots and selects a reduced dimension $r_f$ for the fluid portion of the ROM.
Section~\ref{ssec:specializations} discusses the particular specializations for the AGARD wing that are enabled by the use of block-structured Operator Inference and describes the regularization grid searches performed to identify the optimal regularization levels.
Finally, Sec.~\ref{ssec:performance} analyzes the performance of the block-structured Operator Inference ROM and is the main result of the paper.

\subsection{Data generation}
\label{ssec:data_generation}

\begin{figure}[!htb]
    \centering
    \includegraphics[width=\linewidth]{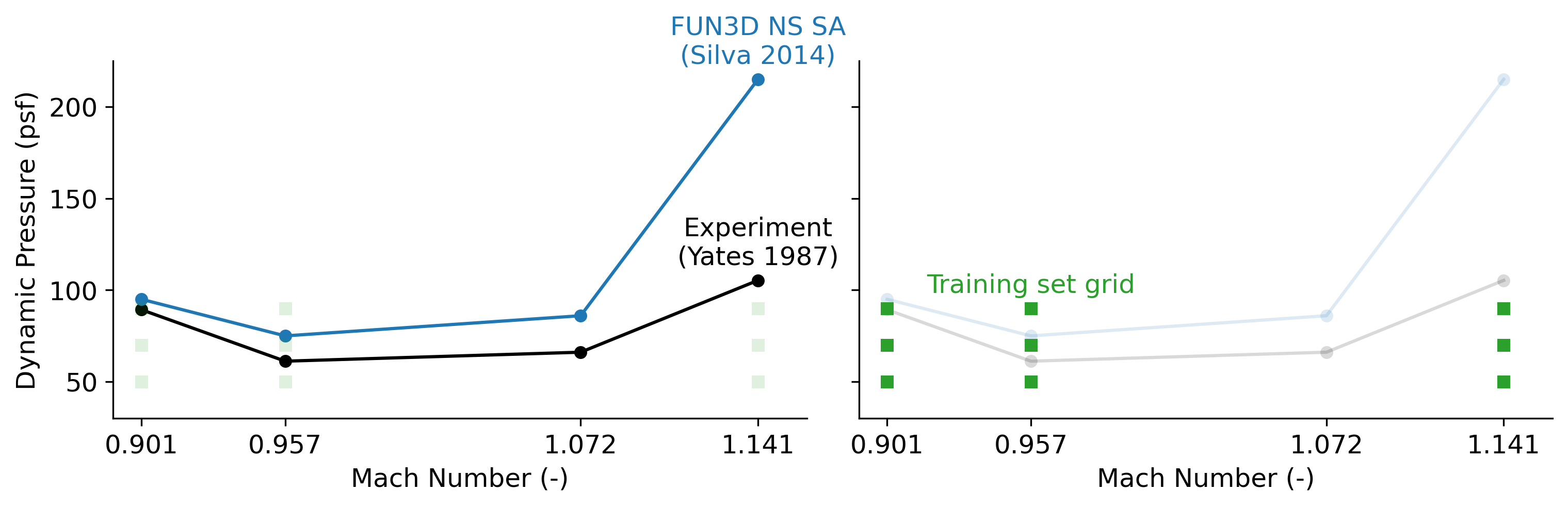}
    \caption{Flutter boundaries from experiments \cite{yates1987agard} and modeling \cite{silva2014agard} in the literature (left) and training set flow conditions (right) for the AGARD 445.6 wing.}
    \label{fig:agard_flutter_boundary}
\end{figure}

To generate training data from the AGARD wing aeroelastic model at varying flow conditions, we select Mach numbers in the set $\mathcal{M}=\{0.901, 0.957, 1.141\}$ and dynamic pressures in the set $\mathcal{Q}=\{50, 70, 90\}$ psf, for a total of nine flow conditions and thus nine training data sets. The Mach numbers match the canonical experimental test conditions from \cite{yates1987agard} and the dynamic pressures provide a uniform grid of values analogous to what might be selected naively for flutter boundary characterization.
We initially included higher dynamic pressures (above 90 psf), but we found that for some flow conditions FUN3D failed to solve at those higher levels. Thus we have limited ourselves to a range of dynamic pressures for which we were able to solve for all flow conditions successfully, allowing us to maintain a uniform grid of training set flow conditions.
The training set flow conditions are plotted in Fig.~\ref{fig:agard_flutter_boundary} along with some experimental and computational flutter points to illustrate the expected behavior at various flow conditions.

The FUN3D full-order model is simulated at each flow condition for $1000$ total timesteps, corresponding to a final time of $t_f = 0.2453$ seconds, thus providing the full data set for use across training and testing. Preliminary testing suggested that this would be a sufficient number of timesteps to provide useful training data and reserve a significant time period from each response for testing of the learned ROMs.
For each flow condition (Mach number and dynamic pressure pair), all four structural generalized velocities are perturbed simultaneously by $gvel_i = 0.1$ while the structural generalized displacements are kept at $gdisp_i = 0$, and then the dynamics are allowed to evolve from that initial perturbation to assess the aeroelastic response.
See Appendix for a detailed description of how the FUN3D input parameters were computed for each flow condition.

\subsection{Fluid data preprocessing}
\label{ssec:preprocessing}

Operator Inference learns ROMs for dynamics with governing equations that have polynomial form and neglects any remaining non-polynomial nonlinear behavior. Although the Navier-Stokes equations do not satisfy this requirement in their primitive state variable form, previous work has shown that transforming the fluid state to specific volume form induces a quadratic form \cite{swischuk2020learning,qian2020lift}. 
We note that this quadratic form applies to the PDE governing equations and not necessarily their discretized form. In particular, the Spalart-Allmaras turbulence model has fourth-order dynamics, and so an intrusively-derived projection-based ROM that perfectly mimicked the FUN3D discretization would include higher order terms. Nonetheless, transforming the snapshots to specific volume variables means that we can target an Operator Inference ROM with quadratic form. We expect that the error incurred due to model misspecification (i.e., ignoring higher order terms) will be less than the approximations introduced by the dimension reduction itself.

FUN3D solves nondimensional governing equations derived from \eqref{eqn:ns_fun3d}
and provides discretized state outputs in primitive form $[\mathbf{p}^\top \ \mathbf{u}^\top \ \mathbf{v}^\top \ \mathbf{w}^\top \ \boldsymbol{\rho}^\top]^\top$, so converting to specific volume variables can be accomplished via a simple postprocessing of the snapshots. This postprocessing replaces the density $\boldsymbol{\rho}$ with the specific volume $\boldsymbol{\xi}$ via the transformation $\boldsymbol{\xi} = 1/\boldsymbol{\rho}$. 
We emphasize that this variable transformation is applied only to the snapshot data, and not to FUN3D itself.
Thus, as shown in \cite{qian2022pdes}, we employ a state vector for Operator Inference of the form $\mathbf{q}_\text{f} = [\mathbf{p}^\top \ \mathbf{u}^\top \ \mathbf{v}^\top \ \mathbf{w}^\top \ \boldsymbol{\xi}^\top]^\top$, and we learn a fluid dynamic ROM with quadratic form.

After transforming the fluid state to induce quadratic form, we preprocess the fluid snapshot data via shifting and scaling to encourage more robust basis identification during POD and inference in the least squares subproblems. We shift the states by subtracting the mean (over time) pressure, velocity, or specific volume value from each pressure, velocity, or specific volume snapshot, respectively. Then we scale the entire fluid snapshot by adjusting each fluid state variable to have pressure and velocity entries in the range of [-1, 1], and specific volume entries in the range of [0, 1]. The full-order model data is now ready for use in the POD and inference steps of the Operator Inference method. Note that shifting the fluid snapshots changes the structure of the ROM ODE. The shifting introduces constant operator terms in addition to the original ROM operators, so the final ROM form consists of constant, linear, and quadratic operators and thus matches the form of \eqref{eqn:rom_block}.

\subsection{Singular value decomposition of fluid snapshots}
\label{ssec:svd}

Now we use the preprocessed training data to identify an appropriate reduced basis for each ROM via the proper orthogonal decomposition (POD). We compute the SVD of the training snapshot matrix for each flow condition, where the fluid training snapshot matrix
$\mathbf{Q}_\text{f} \in \mathbb{R}^{n_\text{f} \times k_\text{train}}$ has $k_\text{train}$ columns, one for each timestep of training data. Here we study $k_\text{train} \in \{100, 300, 500\}$ to investigate the effects of varying the amount of training data required. This is especially important for such a high-dimensional problem because generating each additional training snapshot requires significant computational expense, so we desire that $k_\text{train}$ be as small as possible.

\begin{figure}[!htb]
    \centering
    \includegraphics[width=3.25in]{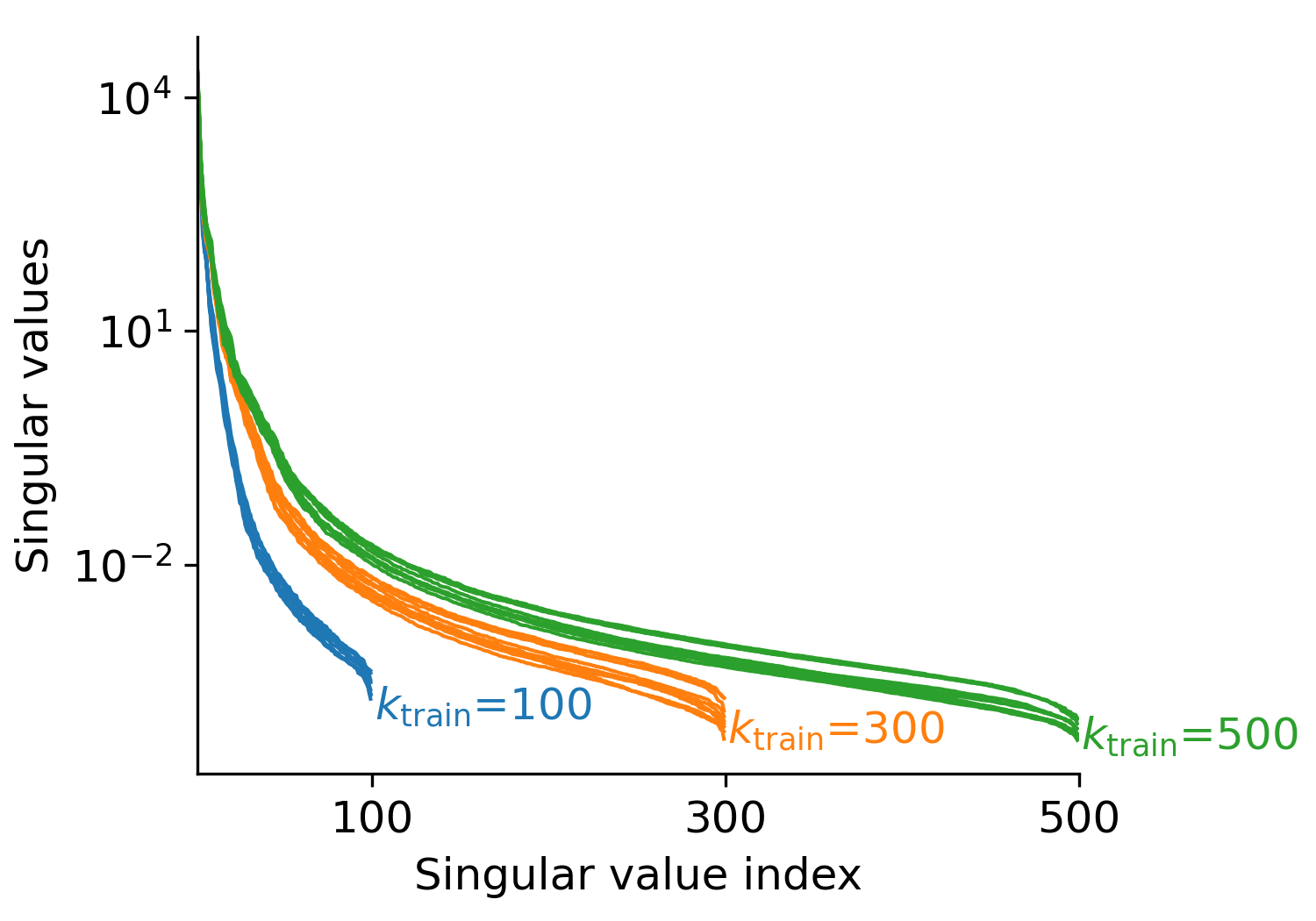}
    \caption{Singular values for snapshot sets of size $k_\text{train} = 100$, $300$ and $500$. For each size, a different snapshot set is analyzed for each of the nine ($M_\infty$, $q_\infty$) flow conditions.}
    \label{fig:singular_values_oneplot}
\end{figure}

Figure \ref{fig:singular_values_oneplot} shows the singular value decay for each flow condition and each studied $k_\text{train}$ value. We can see that for each flow condition, the singular value decay is slower for higher $k_\text{train}$, indicating that there is more information contained in the snapshots.
The singular values decay at similar rates for the first few indices across all $k_\text{train}$ values, but then the $k_\text{train} = 100$ decay rate increases after about five singular values ($r_\text{f} = 5$) and the $k_\text{train} = 300$ decay rate increases after about 25 singular values ($r_\text{f} = 25$). 
While the figure shows that the snapshot set with $k_\text{train} = 500$ contains more information than the set with $k_\text{train} = 300$, the difference is relatively small. We thus choose to use $k_\text{train} = 300$ as the size of our training data set, which provides a good balance between richness of information and magnitude of training data generation costs and storage.

In contrast to the differences in the decay rates between $k_\text{train}$ levels, we see negligible differences between flow conditions (changing Mach number and dynamic pressure). This suggests that similarly sized ROM bases should perform comparably well across the studied space of flow conditions, i.e., we can choose a single $r_\text{f}$ size and use it for all nine ROMs (one ROM per flow condition).

\begin{figure}[!htb]
    \centering
    \includegraphics[width=\linewidth]{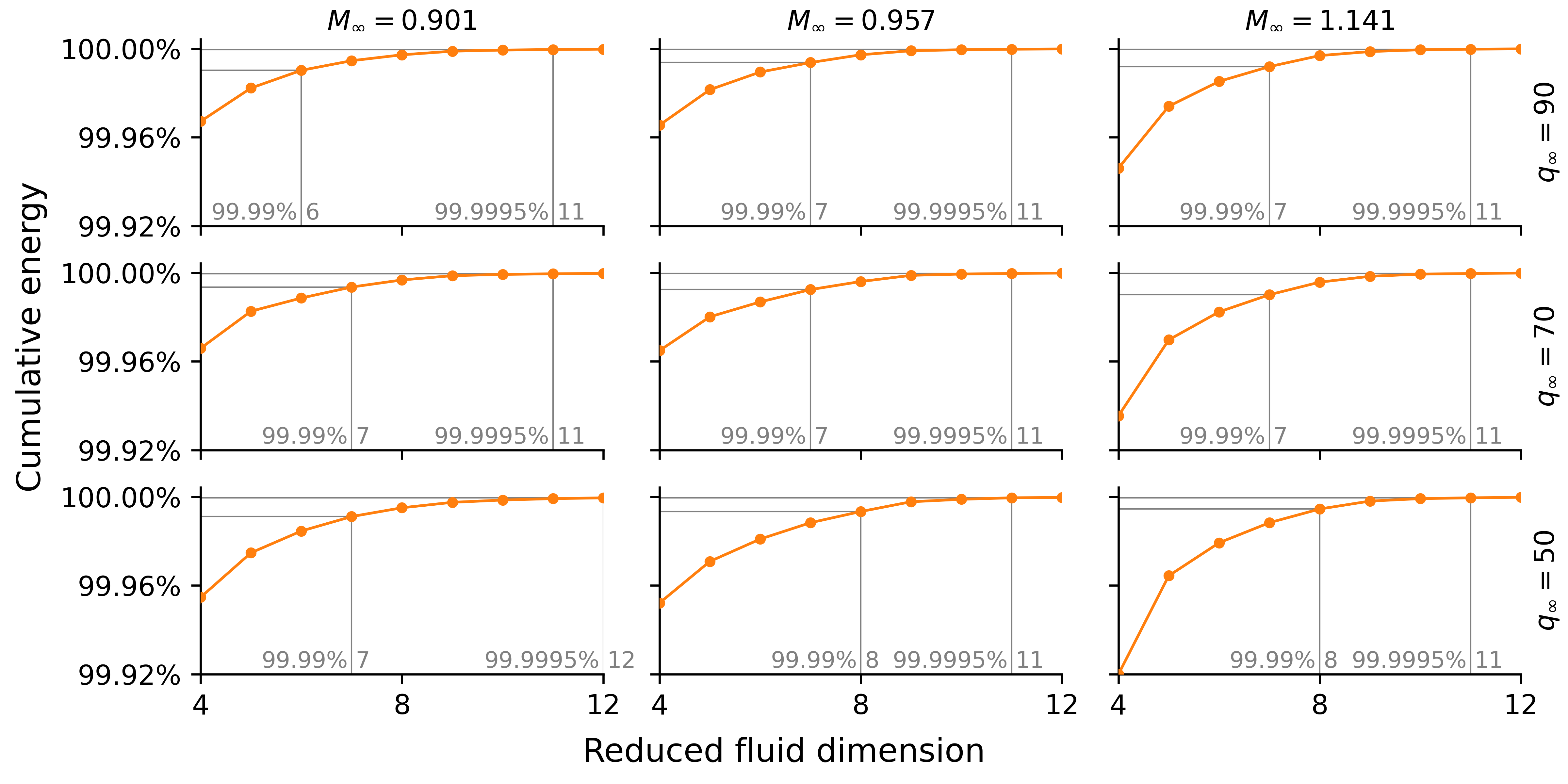}
    \caption{Cumulative energy captured for varying $M_\infty$ and $q_\infty$ where $k_\text{train} = 300$ and thresholds are marked for at least 99.99\% and 99.9995\% cumulative energy captured.}
    \label{fig:cumulative_energy_xrange4to12_k300}
\end{figure}

We also plot the cumulative energy captured for each flow condition for $k_\text{train} = 300$ in Fig.~\ref{fig:cumulative_energy_xrange4to12_k300} as a function of ROM basis size $r_\text{f}$. Each subplot calls out the reduced dimension required to capture 99.99\% and 99.9995\% of the cumulative energy. We see that to capture 99.99\% of the cumulative energy for all nine flow conditions, we require $r_\text{f} = 8$, and to capture 99.9995\% of the cumulative energy for all nine flow conditions, we require $r_\text{f} = 12$; therefore we select these two $r_\text{f}$ levels to generate ROMs for the following ROM performance analyses. These selections will lead to a total coupled ROM dimension of $r = r_\text{s} + r_\text{f} = 8 + 8 = 16$ or $r = 8 + 12 = 20$.

\begin{figure}[!htb]
    \centering
    \includegraphics[width=6.5in]{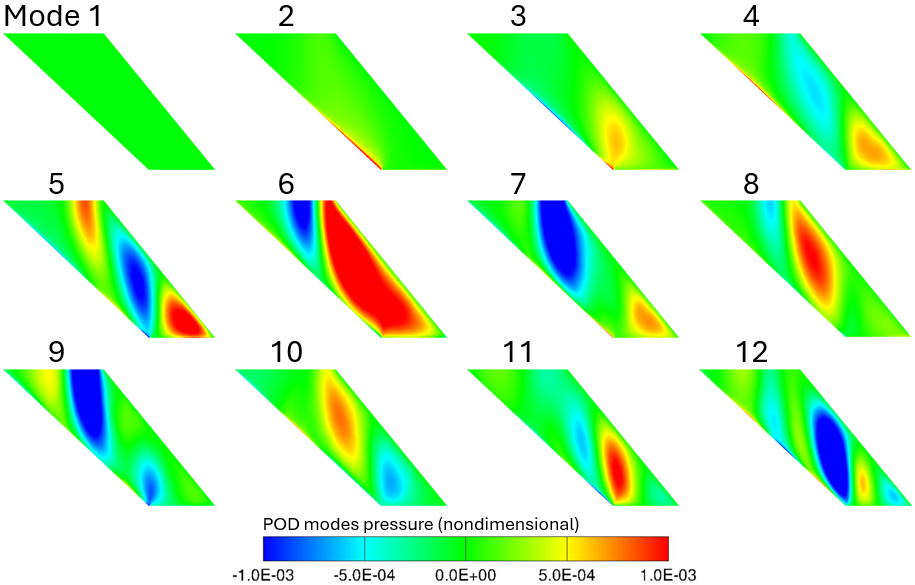}
    \caption{Centered and scaled nondimensional surface pressure portion of each POD mode for $M_\infty = 0.901$, $q_\infty = 50$ psf, and $k_\text{train} = 300$.}
    \label{fig:modes_k300_pressure}
\end{figure}

We use a single reduced basis across all five fluid state variables, so each POD mode contains information about the pressure, velocities, and specific volume over the entire spatial domain of the fluid dynamics. In Fig.~\ref{fig:modes_k300_pressure} we plot the portion of each POD mode that corresponds to the pressure on the surface of the wing for the $M_\infty = 0.901$ and $q_\infty = 50$ psf case with $k_\text{train} = 300$ training snapshots. 
Mode~1 has negligible surface pressure behavior. This is due to the chosen shifting and scaling of snapshots before computing the POD modes as described in Sec.~\ref{ssec:preprocessing}. (Note that in this example, with the chosen scaling the first POD mode predominantly represents variation in specific volume.) Modes~2 and 3 describe localized behavior of the surface pressure near the leading edge and tip of the wing, whereas Modes~4--12 characterize more global behavior across the wing surface.

\subsection{Specializations enabled by the block structure}
\label{ssec:specializations}

Incorporating block structure into Operator Inference provides added flexibility which we exploit to specialize the learning formulation. We use the fluid dynamics and structural dynamics basis matrices to construct a block-structured reduced basis matrix of the form in \eqref{eqn:coupled_basis}. This form has the block-structure-preserving property which permits us to split the monolithic (coupled) least squares problem in \eqref{eqn:lstsq_reg} into the single-physics least squares problems in \eqref{eqn:lstsq_struct} and \eqref{eqn:lstsq_fluid} and take advantage of the block-structured specializations.

Most importantly, we recognize that the structural dynamics governing equation form in \eqref{eqn:fom_ode_struct} is linear while, after lifting to specific volume form and considering the Navier-Stokes governing equations in their PDE form, the fluid dynamics governing equation form is quadratic. The imposition of the linear structural dynamics form can alternatively be seen as intrusively specifying the quadratic structural dynamics operator to be zero, $\widehat{\mathbf{H}}_\text{s} = \mathbf{0}$. We also represent the linear coupling behavior in both the structure-to-fluid and fluid-to-structure directions. Referring back to the derivation of the block structure from Sec.~\ref{sssec:coupled_fom}, this means that we can take advantage of the block structure of the coupling terms to impose that the bilinear and quadratic coupling operators should be zero as well, i.e., $\widehat{\mathbf{L}}_\text{s} = \widehat{\mathbf{G}}_\text{s} = \widehat{\mathbf{G}}_\text{f} = \widehat{\mathbf{L}}_\text{f} = \mathbf{0}$.

These specializations lead us to update the desired ROM form that we originally specified in \eqref{eqn:rom_block} by removing the five zeroed-out blocks of the quadratic operator and replacing them with zero blocks. We can now write our target ROM dynamics in the form
\begin{equation}
    \label{eqn:rom_block_zeros_intrusive}
    \begin{bmatrix}\dot{\widehat{\mathbf{q}}}_\text{s} \\ \dot{\widehat{\mathbf{q}}}_\text{f}\end{bmatrix}
    =
    \begin{bmatrix}\widehat{\mathbf{c}}_\text{s} \\ \widehat{\mathbf{c}}_\text{f}\end{bmatrix}
    +
    \begin{bmatrix}\widehat{\mathbf{A}}_\text{s} & \widehat{\mathbf{E}}_\text{s} \\ \widehat{\mathbf{E}}_\text{f} & \widehat{\mathbf{A}}_\text{f}\end{bmatrix}
    \begin{bmatrix}\widehat{\mathbf{q}}_\text{s} \\ \widehat{\mathbf{q}}_\text{f}\end{bmatrix}
    +
    \begin{bmatrix}\mathbf{0} & \mathbf{0} & \mathbf{0} \\ \mathbf{0} & \mathbf{0} & \widehat{\mathbf{H}}_\text{f}\end{bmatrix}
    \begin{bmatrix}\left(\widehat{\mathbf{q}}_\text{s} \otimes \widehat{\mathbf{q}}_\text{s}\right) \\ \left(\widehat{\mathbf{q}}_\text{s} \otimes \widehat{\mathbf{q}}_\text{f}\right) \\ \left(\widehat{\mathbf{q}}_\text{f} \otimes \widehat{\mathbf{q}}_\text{f}\right)\end{bmatrix}.
\end{equation}

Crucially, this new form leads to many simplifications in the least squares subproblems from \eqref{eqn:lstsq_struct} and \eqref{eqn:lstsq_fluid}. We simplify the least squares input and output matrices $\widehat{\mathbf{D}}_\text{s}$, $\widehat{\mathbf{R}}_\text{s}$, $\widehat{\mathbf{O}}_\text{s}$, $\widehat{\mathbf{D}}_\text{f}$, $\widehat{\mathbf{R}}_\text{f}$, and $\widehat{\mathbf{O}}_\text{f}$ from Sec.~\ref{ssec:block_opinf} by removing the intrusively specified blocks and shifting any known terms to the right-hand-side matrices. However, the intrusively specified blocks in this problem are all zeros, so we just remove them from the computations. Then the structural dynamics least squares problem matrices are
\begin{align}
    \widehat{\mathbf{D}}_\text{s} &= \begin{bmatrix} \mathbf{1}_k^\top & \widehat{\mathbf{Q}}_\text{s}^\top & \widehat{\mathbf{Q}}_\text{f}^\top \end{bmatrix} \\
    \widehat{\mathbf{O}}_\text{s} &= \begin{bmatrix}\widehat{\mathbf{c}}_\text{s} & \widehat{\mathbf{A}}_\text{s} & \widehat{\mathbf{E}}_\text{s} \end{bmatrix} \\
    \widehat{\mathbf{R}}_\text{s} &= \begin{bmatrix}\dot{\widehat{\mathbf{Q}}}_\text{s}\end{bmatrix}
\end{align}
and the fluid dynamics least squares problem matrices are
\begin{align}
    \widehat{\mathbf{D}}_\text{f} &= \begin{bmatrix} \mathbf{1}_k^\top & \widehat{\mathbf{Q}}_\text{s}^\top & \widehat{\mathbf{Q}}_\text{f}^\top & \left(\widehat{\mathbf{Q}}_\text{f} \otimes \widehat{\mathbf{Q}}_\text{f}\right)^\top\end{bmatrix} \\
    \widehat{\mathbf{O}}_\text{f} &= \begin{bmatrix}\widehat{\mathbf{c}}_\text{f} & \widehat{\mathbf{E}}_\text{f} & \widehat{\mathbf{A}}_\text{f} & \widehat{\mathbf{H}}_\text{f}\end{bmatrix} \\
    \widehat{\mathbf{R}}_\text{f} &= \begin{bmatrix}\dot{\widehat{\mathbf{Q}}}_\text{f}\end{bmatrix}.
\end{align}
Finally, the separation of the learning step into two least squares subproblems permits us to separately regularize each block in the operator matrices $\widehat{\mathbf{O}}_\text{s}$ and $\widehat{\mathbf{O}}_\text{f}$.
In principle, we could choose independent regularization parameters for each operator $\widehat{\mathbf{c}}_\text{s}$, $\widehat{\mathbf{A}}_\text{s}$, $\widehat{\mathbf{E}}_\text{s}$, $\widehat{\mathbf{c}}_\text{f}$, $\widehat{\mathbf{A}}_\text{f}$, $\widehat{\mathbf{E}}_\text{f}$, and $\widehat{\mathbf{H}}_\text{f}$, analogously to the form of the independent regularization parameters in \eqref{eqn:lstsq_reg}, but it would be computationally challenging to determine optimal values for the resulting seven regularization parameters. To mitigate this complexity, in our problem setup we choose to
regularize the operators $\widehat{\mathbf{c}}_\text{s}$, $\widehat{\mathbf{A}}_\text{s}$ and $\widehat{\mathbf{E}}_\text{s}$ together
so that the structural subproblem's regularization has the form
\begin{equation*}
    \mathcal{R}_\text{s}(\widehat{\mathbf{O}}_\text{s}) = \gamma_\text{s}^\text{linear} \left(\left\|\widehat{\mathbf{c}}_\text{s}\right\|_F^2 + \left\|\widehat{\mathbf{A}}_\text{s}\right\|_F^2 + \left\|\widehat{\mathbf{E}}_\text{s}\right\|_F^2\right)
\end{equation*}
and we choose to 
regularize the operators $\widehat{\mathbf{c}}_\text{f}$, $\widehat{\mathbf{A}}_\text{f}$ and $\widehat{\mathbf{E}}_\text{f}$ together,
so that the fluid subproblem's regularization has the form
%%%
\begin{equation*}
    \mathcal{R}_\text{f}(\widehat{\mathbf{O}}_\text{f}) = \gamma_\text{f}^\text{linear} \left(\left\|\widehat{\mathbf{c}}_\text{f}\right\|_F^2 + \left\|\widehat{\mathbf{A}}_\text{f}\right\|_F^2 + \left\|\widehat{\mathbf{E}}_\text{f}\right\|_F^2\right) + \gamma_\text{f}^\text{quadratic} \left\|\widehat{\mathbf{H}}_\text{f}\right\|_F^2.
\end{equation*}
This choice leads to a three-dimensional regularization hyperparameter space, where we need to select the three regularization parameters $\gamma_\text{s}^\text{linear}$, $\gamma_\text{f}^\text{linear}$, and $\gamma_\text{f}^\text{quadratic}$.
We perform a coarse, logarithmically-spaced grid search followed by a more targeted, linearly-spaced grid search to identify the optimal regularization values for the three hyperparameters. During these grid searches, we also incorporate the previously mentioned bounded growth constraint, which we specify to be $\alpha = 10$ after a linear grid search.

\subsection{Performance of block-structured Operator Inference}
\label{ssec:performance}

This section compares the performance of the block-structured and monolithic Operator Inference ROMs for the AGARD wing by using the full-order model's predictions and performance as a reference.
Section \ref{sssec:accuracy} investigates the accuracy of each ROM type, and then Section \ref{sssec:complexity} investigates the computational complexity and associated prediction costs of each ROM type.

\subsubsection{Accuracy}
\label{sssec:accuracy}

After identifying the appropriate regularization levels for each flow condition, we proceed to analyzing the results of the block-structured Operator Inference ROM. Our quantities of interest for the AGARD wing are the lift coefficient $C_L$ for the fluid dynamics and the generalized displacements $gdisp_i$ for the structural dynamics. We choose these quantities because the lift coefficient represents an integrated effect of the fluid dynamics on the structural dynamics and the 
generalized displacements can be used to assess the stability of the coupled dynamics for flutter boundary characterization \cite{jacobson2019mpm}. We note that the aerodynamic accuracy of the ROMs could be further assessed by analyzing generalized aerodynamic forces. Such an analysis could give additional insight into the conditions where ROM predictions deviate from the corresponding high-fidelity CFD result.

\begin{figure}[!htb]
    \centering
    \includegraphics[width=0.95\linewidth]{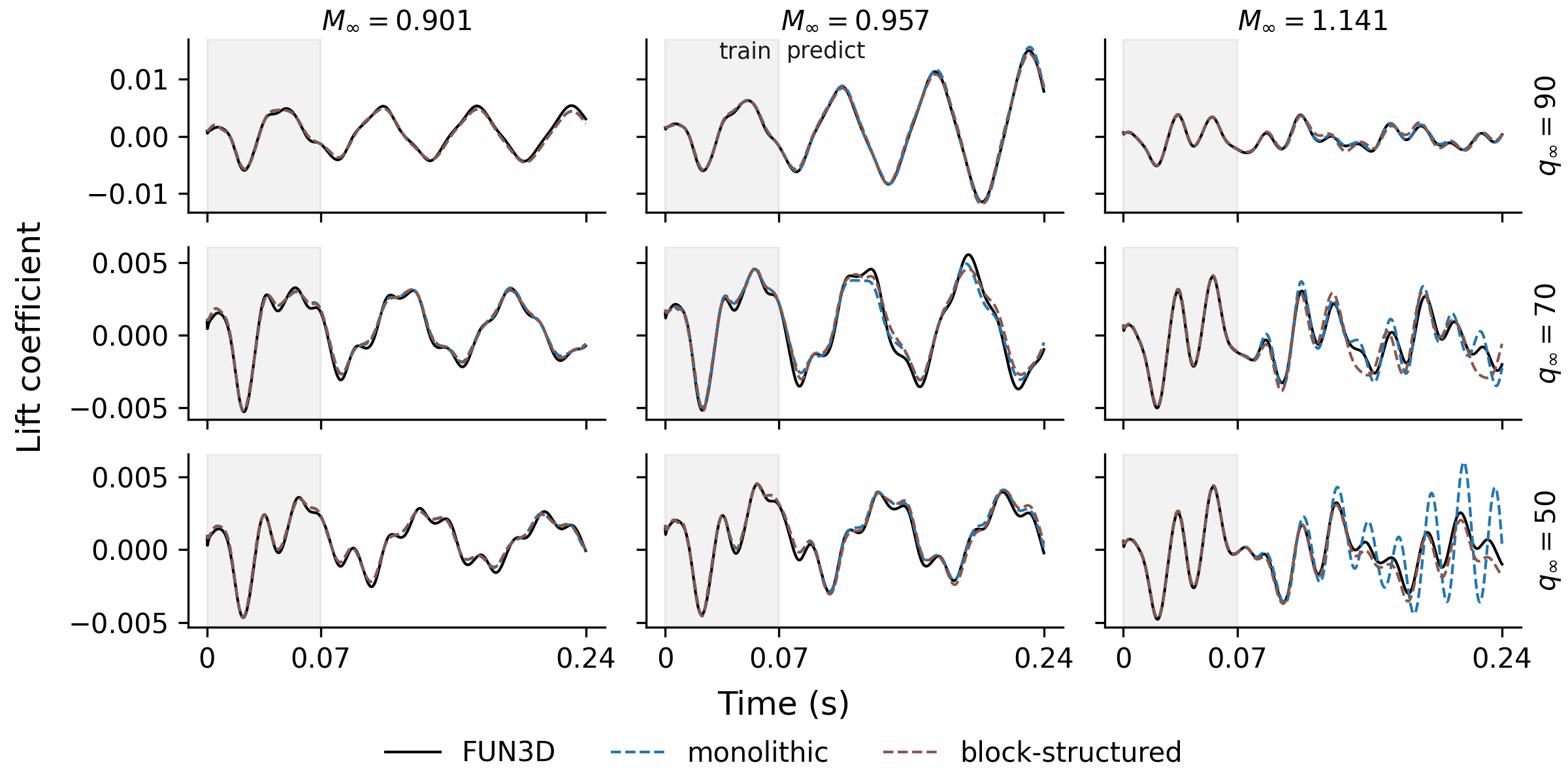}
    \caption{Lift coefficient ($C_L$) predictions for monolithic and block-structured Operator Inference with $k_\text{train} = 300$ and $r_\text{f} = 8$.}
    \label{fig:CL_k300_r8}
\end{figure}

\begin{figure}[!htb]
    \centering
    \includegraphics[width=0.95\linewidth]{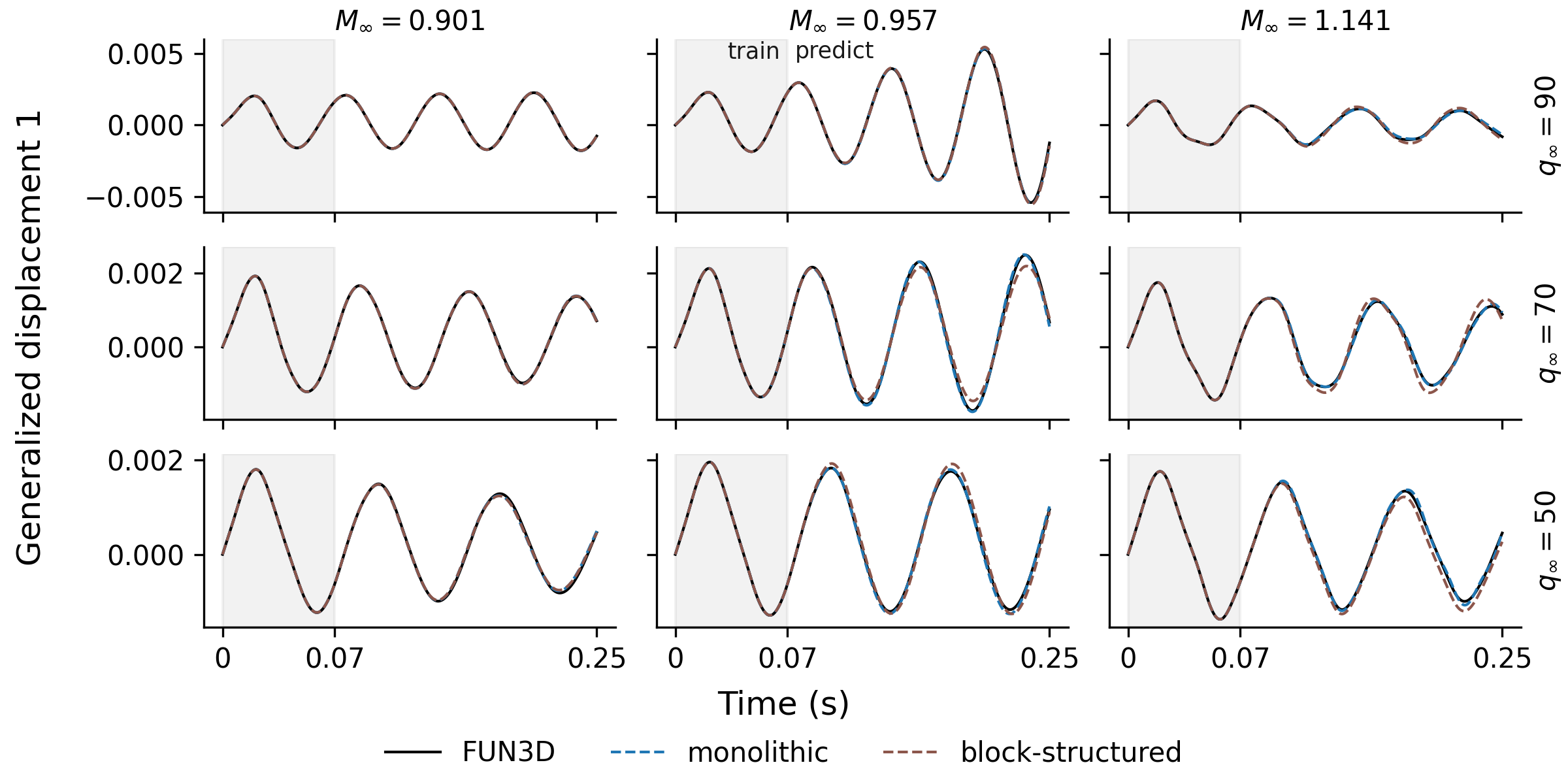}
    \caption{First generalized displacement ($gdisp_1$) predictions for monolithic and block-structured Operator Inference with $k_\text{train} = 300$ and $r_\text{f} = 8$.}
    \label{fig:gdisp1_k300_r8}
\end{figure}

\begin{figure}[!htb]
    \centering
    \includegraphics[width=0.95\linewidth]{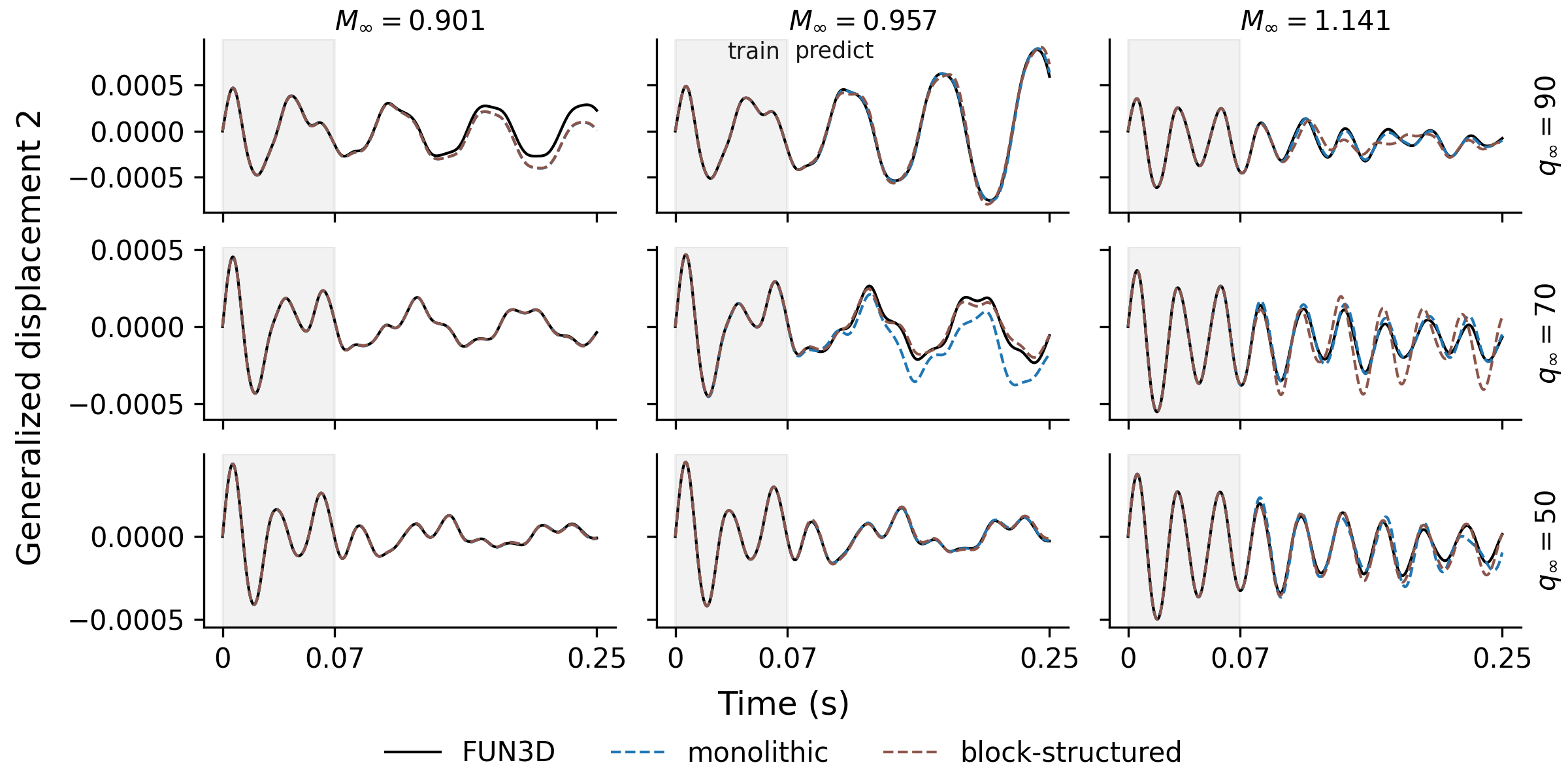}
    \caption{Second generalized displacement ($gdisp_2$) predictions for monolithic and block-structured Operator Inference with $k_\text{train} = 300$ and $r_\text{f} = 8$.}
    \label{fig:gdisp2_k300_r8}
\end{figure}

\begin{figure}[!htb]
    \centering
    \includegraphics[width=0.95\linewidth]{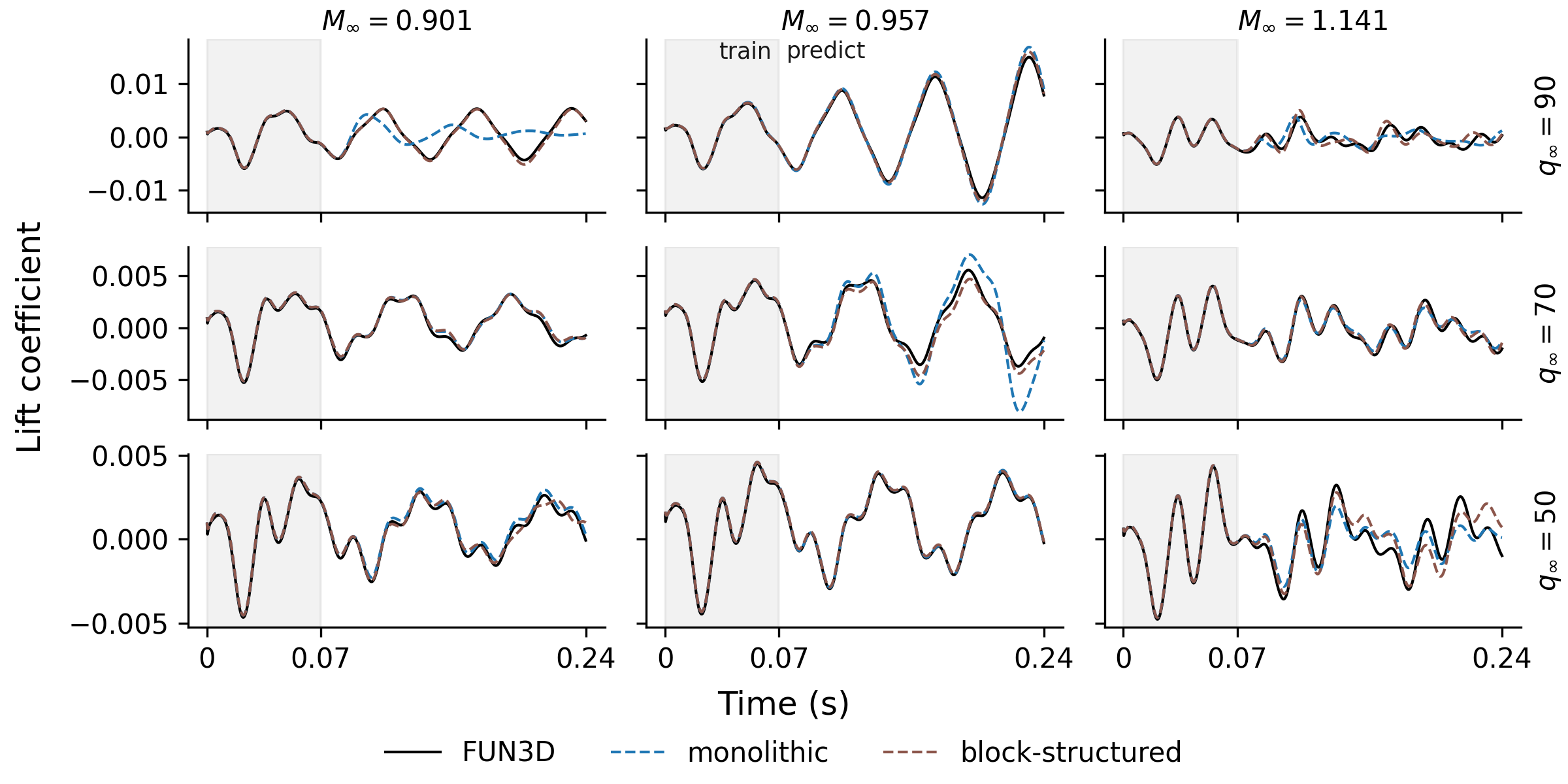}
    \caption{Lift coefficient ($C_L$) predictions for monolithic and block-structured Operator Inference with $k_\text{train} = 300$ and $r_\text{f} = 12$.}
    \label{fig:CL_k300_r12}
\end{figure}

\begin{figure}[!htb]
    \centering
    \includegraphics[width=0.95\linewidth]{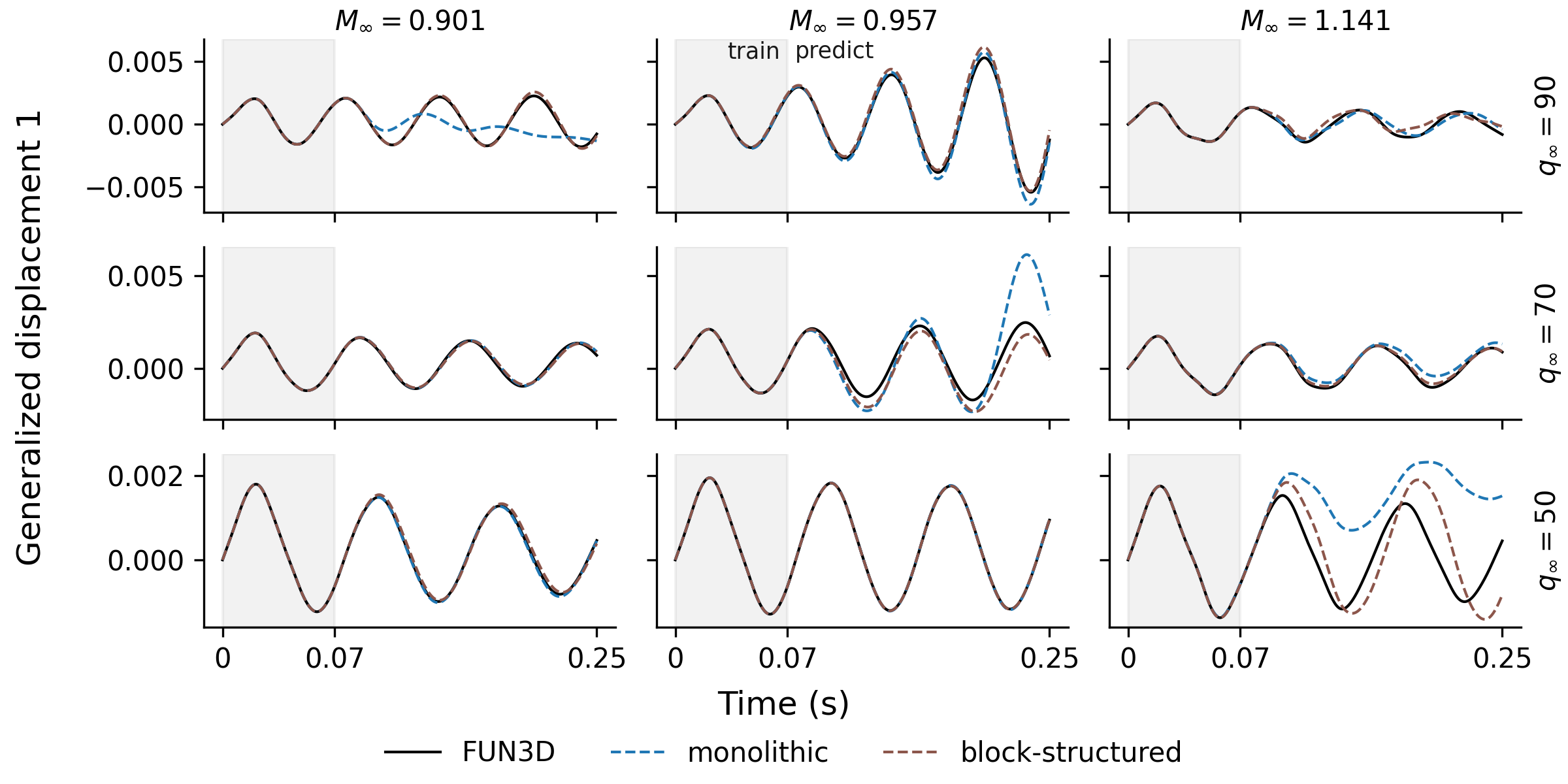}
    \caption{First generalized displacement ($gdisp_1$) predictions for monolithic and block-structured Operator Inference with $k_\text{train} = 300$ and $r_\text{f} = 12$.}
    \label{fig:gdisp1_k300_r12}
\end{figure}

\begin{figure}[!htb]
    \centering
    \includegraphics[width=0.95\linewidth]{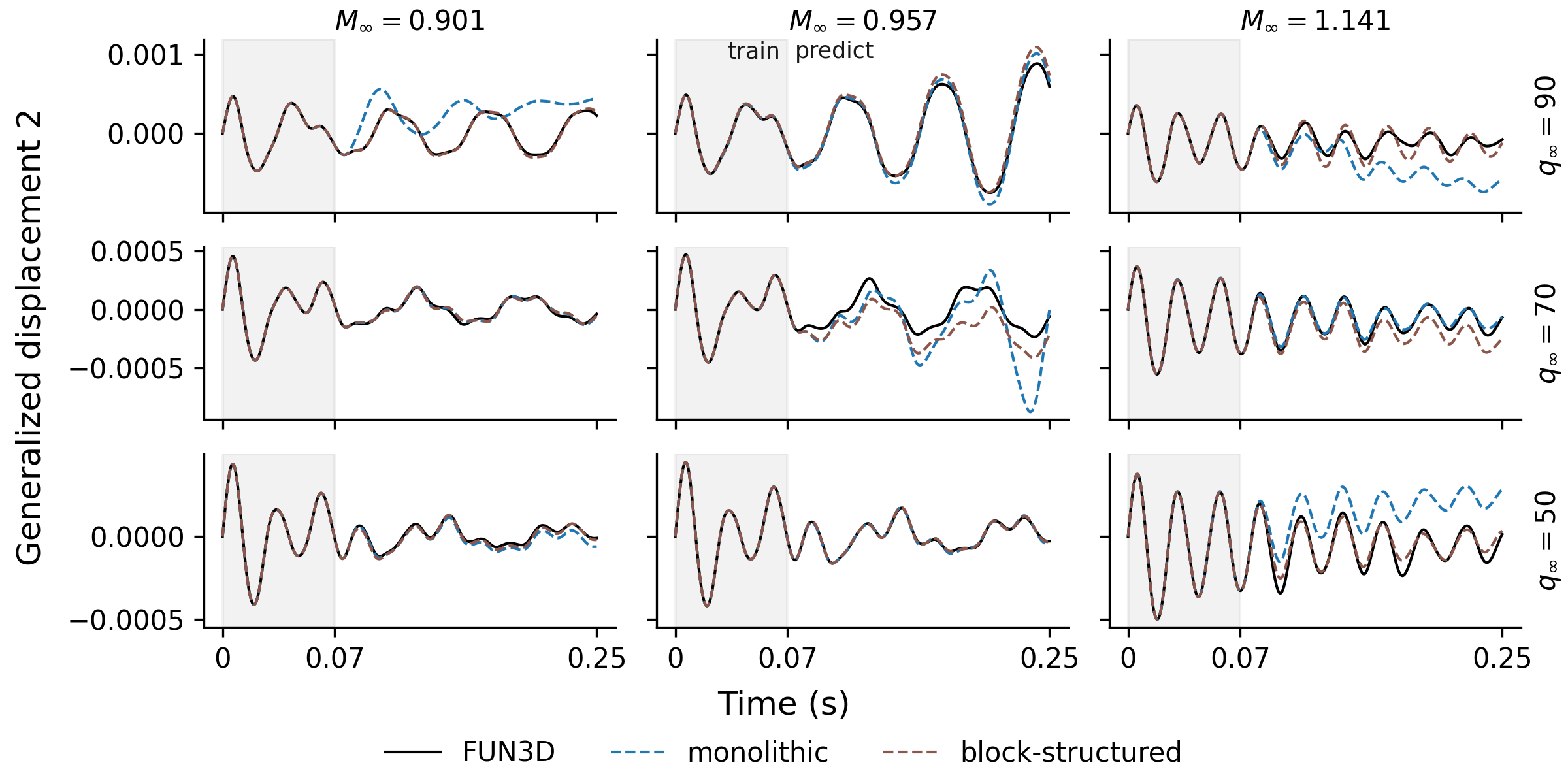}
    \caption{Second generalized displacement ($gdisp_2$) predictions for monolithic and block-structured Operator Inference with $k_\text{train} = 300$ and $r_\text{f} = 12$.}
    \label{fig:gdisp2_k300_r12}
\end{figure}

We use the ROMs to predict forward in time to the end of the testing regime ($t_f=0.2453$, i.e., 1000 timesteps) and compare to the full-order model predictions for each flow condition. For each ROM (learned separately for each of the nine flow conditions), we perturb the aeroelastic system in the same manner as during the training data generation in Sec. \ref{ssec:data_generation} by setting all four generalized velocities to a nonzero initial value of $gvel_i = 0.1$ and then investigating the dynamics that occur due to the initial perturbations. Figs.~\ref{fig:CL_k300_r8}--\ref{fig:gdisp2_k300_r12} show the improved accuracy of the block-structured Operator Inference predictions compared to monolithic Operator Inference for the lift coefficient, $C_L$, and the first two generalized displacements, $gdisp_1$ and $gdisp_2$ for $r_\text{f} = 8, 12$. We see that the $M_\infty = 0.901$, $q_\infty = 50, 70$ psf flow conditions are accurately predicted by both methods, which is not surprising because these are the least complex flow conditions. The higher $q_\infty$ flow conditions, which approach the flutter boundary, are more clearly improved by the block-structured method for some quantities, but not for all of them. The supersonic $M_\infty = 1.141$ flow conditions are not perfectly predicted by either method, but are clearly more stable when using the block-structured method.

\begin{figure}[!htb]
    \centering
    \includegraphics[width=0.85\linewidth]{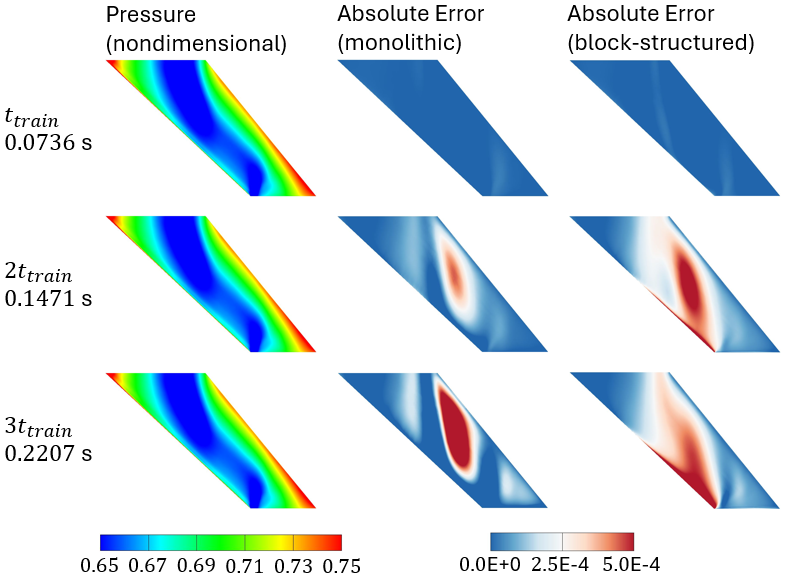}
    \caption{Pressure and pressure absolute error wing surface contours for $M_\infty=0.957$, $q_\infty=70$ psf, $k_\text{train}=300$, and $r_\text{f}=8$.}
    \label{fig:pressure_contours_k300_r8}
\end{figure}

\begin{figure}[!htb]
    \centering
    \includegraphics[width=0.85\linewidth]{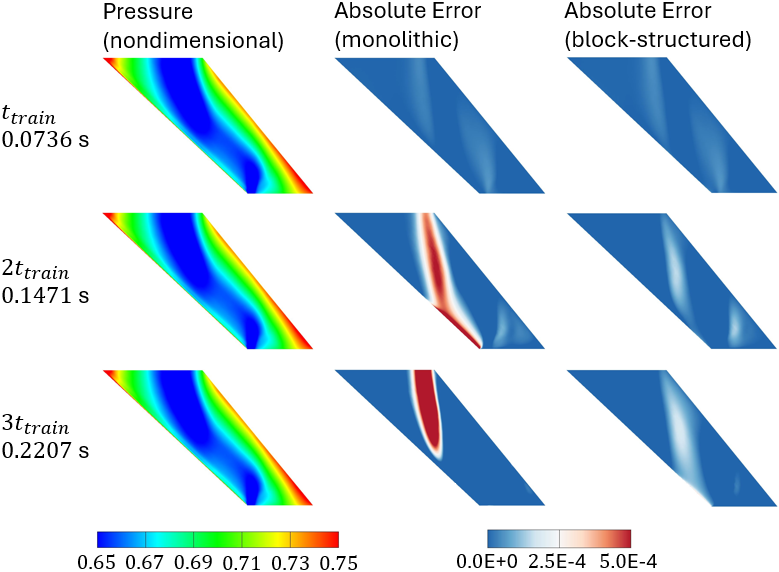}
    \caption{Pressure and pressure absolute error wing surface contours for $M_\infty=0.957$, $q_\infty=70$ psf, $k_\text{train}=300$, and $r_\text{f}=12$.}
    \label{fig:pressure_contours_k300_r12}
\end{figure}

\clearpage
The accuracy of pressure distribution prediction on the surface of the wing provides more insight into the ROM's performance, especially because the surface pressure distribution is the source of the lift coefficient calculation.
Figures~\ref{fig:pressure_contours_k300_r8} and \ref{fig:pressure_contours_k300_r12} illustrate the surface pressure distribution and absolute pressure error on the surface of the AGARD wing at three different timesteps corresponding to the end of the training period, $t_\text{train} = t(k_\text{train})$, and at two and three times the training period, $2t_\text{train}$ and $3t_\text{train}$. The magnitude of these errors is small due to the small perturbations that we use to excite the dynamics of the system. However, for computations like flutter prediction this is acceptable since we are primarily interested in the stability of the coupled dynamics. 
We can see that the largest pressure prediction differences occur across the location where the main shock discontinuity is beginning to form. Additionally, the largest errors are along the leading edge of the wing and just upstream and downstream of the shock discontinuity. POD modes are known to struggle with representing discontinuous changes in the state, so this is not surprising.

We compare these error contours to the $C_L$ time histories in Figs.~\ref{fig:CL_k300_r8} and \ref{fig:CL_k300_r12}. We see that the poor performance of the monolithic ROM for $r_\text{f}=12$ in the contour plots is because the monolithic $C_L$ approximation does not track the full-order model $C_L$ for that flow condition. Conversely, we see that both ROMs have small $C_L$ errors for $r_\text{f}=8$ (Fig.~\ref{fig:CL_k300_r8}),
and the fact that the block-structured ROM has larger error is a function of the specific timesteps at which we plotted the contours.

Inspecting Figs.~\ref{fig:CL_k300_r8}--\ref{fig:gdisp2_k300_r12} reveals tradeoffs between accurately predicting the $C_L$ response and the generalized displacement responses, along with related tradeoffs in accuracy from using different reduced fluid dimensions ($r_\text{f} = 8, 12$). We can see that $C_L$ and $gdisp_1$ are predicted more accurately for $r_\text{f} = 8$, but $gdisp_2$ is predicted more accurately for $r_\text{f} = 12$.
As is often the case in POD-based ROMs, the lower-dimension ROM ($r_\text{f}=8$) does an excellent job of capturing the response of lower frequency dynamic behavior (i.e., the first generalized displacement), but does not fully resolve the higher frequency behavior of the second generalized displacement.
Increasing the dimension of the POD basis guarantees that the training snapshot data can be represented with lower error, but it does not guarantee more accurate ROM predictions, especially for predictions beyond the end of the training period.
For example, 
we see that the $gdisp_1$ predictions for $r_\text{f}=8$ in Fig.~\ref{fig:gdisp1_k300_r8} are more accurate than the equivalent predictions for $r_\text{f}=12$ in Fig.~\ref{fig:gdisp1_k300_r12}. 
This is likely because the $r_\text{f} = 12$ ROM overfits to the training data when we increase the reduced dimension, which is a common tradeoff that is seen across data-driven modeling. This hypothesis is supported by the better (more stable) predictions of the block-structured ROM compared to the monolithic ROM for $r_\text{f} = 12$, recalling that a benefit of the block-structured approach is the reduced data requirement due to the lower number of operator entries that must be learned (see Sec.~\ref{sssec:complexity} for further discussion of this distinction).

To check the accuracy of the ROMs at other $r_\text{f}$ levels, we use a relative root-mean-square-error (RMSE) metric normalized by the range of the full-order model-predicted quantity of interest over the prediction time period. Here our quantity of interest is the lift coefficient $C_L$ that was plotted in Figs.~\ref{fig:CL_k300_r8} and \ref{fig:CL_k300_r12} above.

We define the lift coefficient error metric as
\begin{equation} \label{eqn:error_metric}
    \varepsilon_\text{rel}(\mathbf{C_L}) = \frac
    {\sqrt{ \frac{1}{k_\text{predict}} \sum_{i=1}^{k_\text{predict}} (C_{L,i}^\text{FOM}-C_{L,i}^\text{ROM})^2}}
    {\max(\mathbf{C_L}^\text{FOM}) - \min(\mathbf{C_L}^\text{FOM})}
\end{equation}
where $\mathbf{C_L}^\text{FOM} \in \mathbb{R}^{k_\text{predict}}$ and $\mathbf{C_L}^\text{ROM} \in \mathbb{R}^{k_\text{predict}}$ are the full-order model-predicted and ROM-predicted lift coefficients over all timesteps in the prediction regime
$[t_\text{train}, \ t_\text{final}] = [0.0736, \ 0.2453]$
seconds and where
$k_\text{predict}$ is the number of timesteps in the prediction regime.
Across the flow conditions, $C_L$ ranges from 0.005391 to 0.026499, thus providing a physically meaningful normalization value for each flow condition to assess the magnitude of the ROM prediction errors.

\begin{figure}[!htb]
    \centering
    \includegraphics[width=0.95\linewidth]{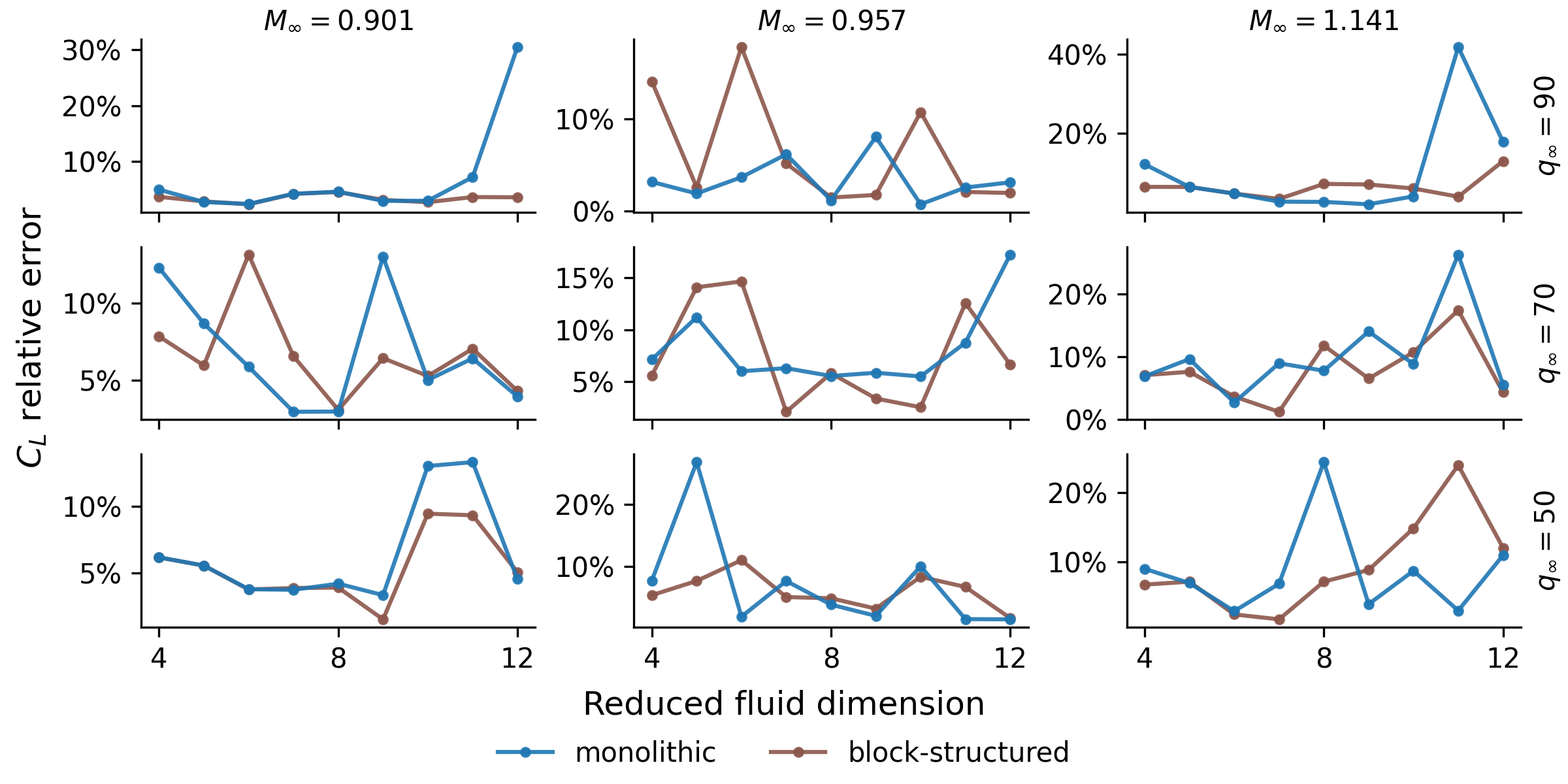}
    \caption{Time-integrated lift coefficient relative prediction error comparing monolithic and block-structured Operator Inference with $k_\text{train} = 300$.
    }
    \label{fig:CL_rel_err_global_k300}
\end{figure}

Figure \ref{fig:CL_rel_err_global_k300} shows the error metric $\varepsilon_\text{rel}(\mathbf{C_L})$ in \eqref{eqn:error_metric} for the lift coefficient under varying reduced fluid dimension and all nine flow conditions.
We see that there is not a clear relationship between increasing reduced fluid dimension and decreasing relative error. Thus we should focus on choosing a reduced fluid dimension such that the ROM is as stable as possible, especially since Fig.~\ref{fig:CL_rel_err_global_k300} illustrates that there are some choices of $r_\text{f}$ that lead to more unstable (inaccurate) ROMs. Overall, we see that there is not an especially clear pattern of one method outperforming the other, but the block-structured method tends to match the monolithic Operator Inference performance, with each method having a few poor-performing data points, while the block-structured method uses a significantly smaller ROM (with a faster prediction time), as we discuss in Sec.~\ref{sssec:complexity}.

\subsubsection{Computational complexity}
\label{sssec:complexity}

\begin{figure}[htb]
    \centering
    \begin{subfigure}[b]{0.56\textwidth}
        \centering
        \includegraphics[width=\textwidth]{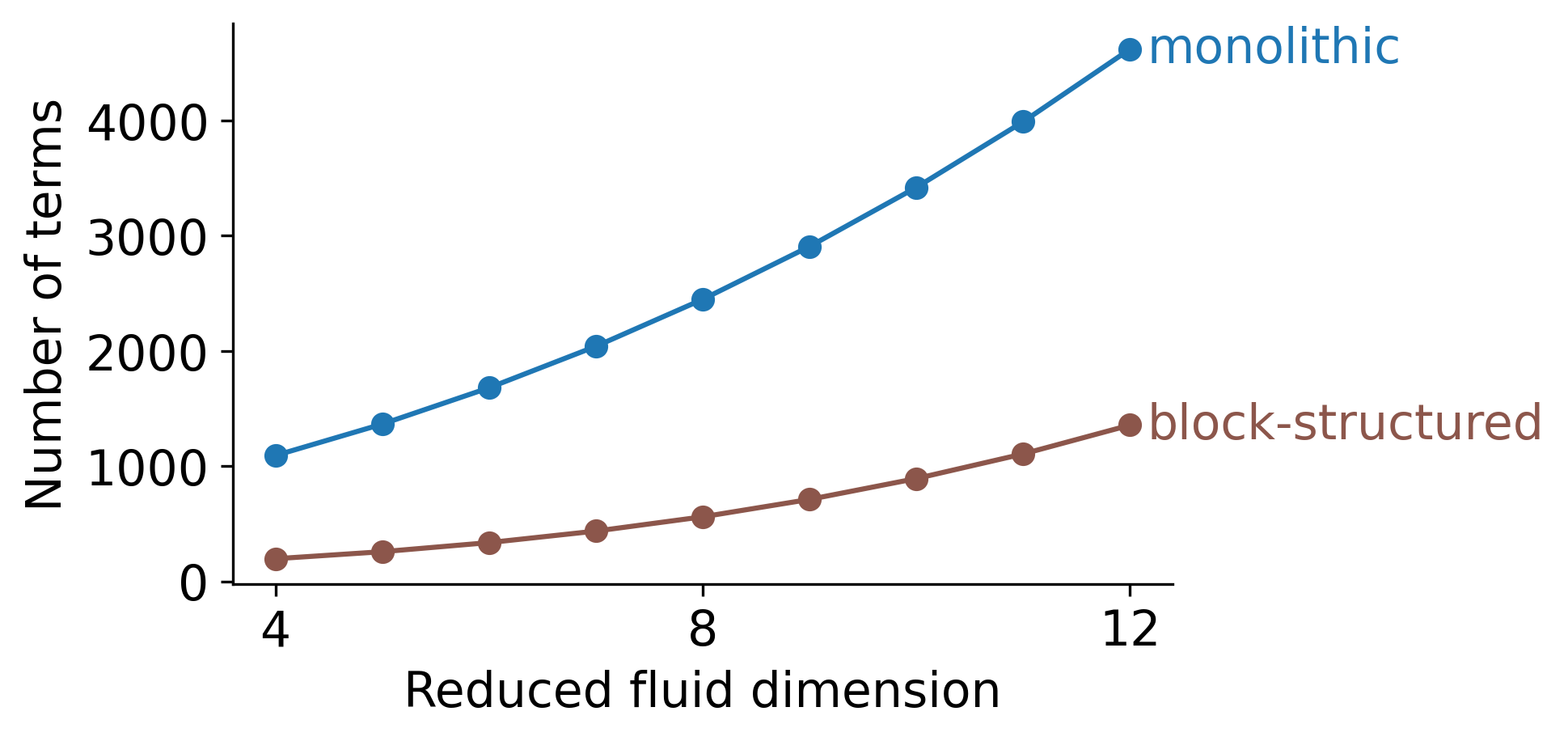}
        \caption{Number of terms to learn}
        \label{fig:num_terms}
    \end{subfigure}
    \begin{subfigure}[b]{0.43\textwidth}
        \centering
        \includegraphics[width=\textwidth]{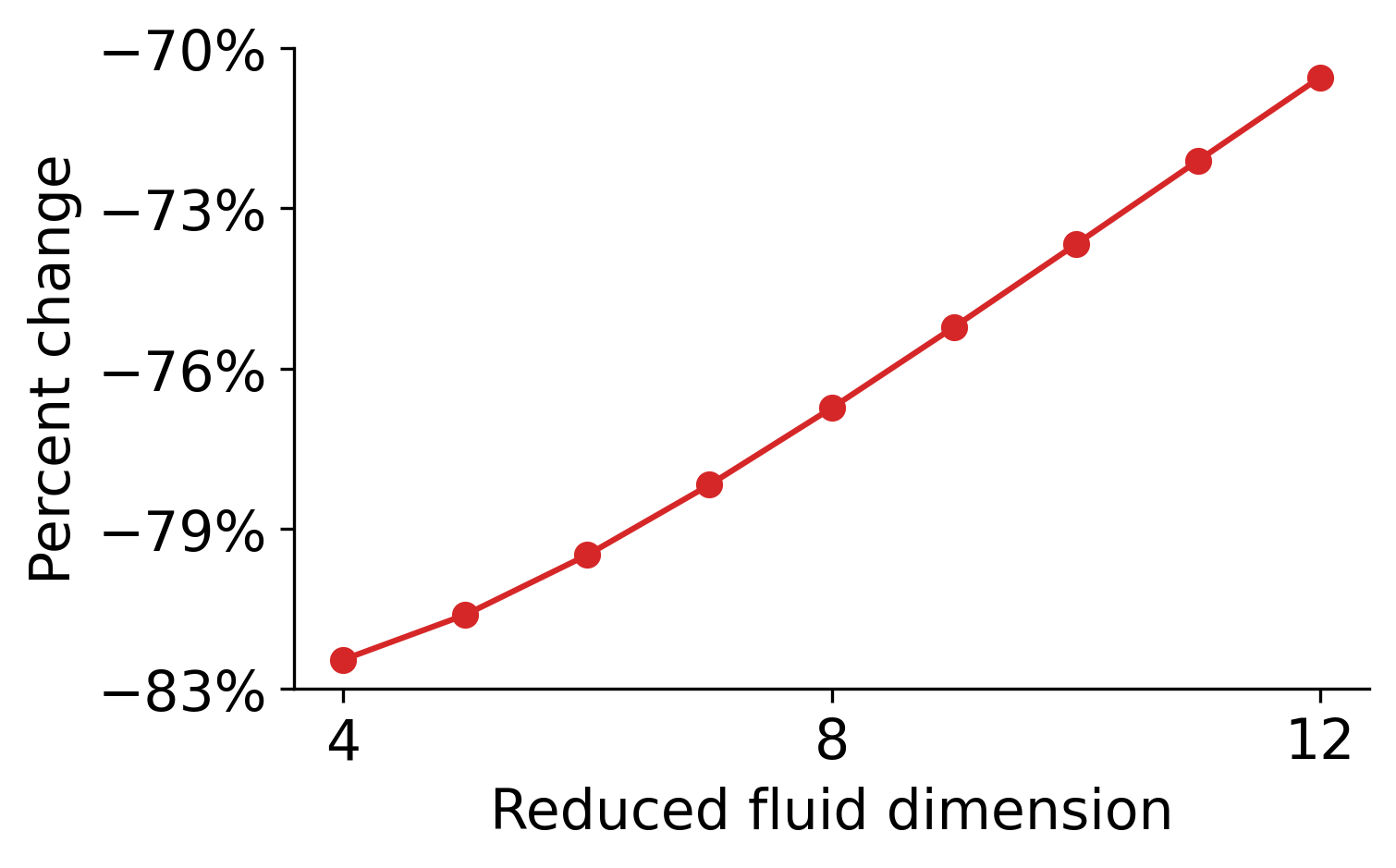}
        \caption{Percent change in number of terms to learn}
        \label{fig:percent_change}
    \end{subfigure}
    \caption{Computational complexity of the aeroelastic AGARD Operator Inference problem for varying reduced fluid dimension.}
    \label{fig:complexity}
\end{figure}

A significant benefit of the block-structured Operator Inference method is the reduced computational expense of the learning and prediction steps. This savings is due to the smaller number of nonzero operator entries that must be learned in the least squares step and computed during each temporal propagation step. Figure \ref{fig:complexity} illustrates these savings for a constant $r_s=8$ and varying $r_f$. We see that for small $r_f$, the percent reduction in the number of entries to learn is large, but the actual decrease in the number of entries is small. Conversely, for large $r_f$, the percent reduction gets smaller but the actual decrease in the number of entries grows with the increase in $r_f$. Of course, this relationship is specific to the block-structure of the aeroelastic problem studied in this paper, and different savings trends would be observed for dynamics with different mathematical structure.

For small reduced fluid dimension ($r_\text{f}$), the difference in computational complexity between the ROMs is mostly due to the unnecessary inclusion of the quadratic structural operator $\widehat{\mathbf{H}}_\text{s}$ in the monolithic ROM, since the linear, bilinear, and quadratic coupling operators have relatively few terms.
As $r_\text{f}$ increases, the complexity of the fluid dynamics operators and the coupling operators begins to dominate in both ROMs. However, some of these operators are then set to zero for the block-structured ROM, thus illustrating the reduced complexity of block-structured Operator Inference. This appears in Fig.~\ref{fig:num_terms} as the increasing gap between the two lines, leading to the behavior in Fig.~\ref{fig:percent_change}. To summarize, the complexity difference is due entirely to the imposition of zero blocks in the block-structured ROM, but the magnitude of the difference is dependent on the reduced fluid dimension, at least for the AGARD example. In other model reduction applications, intrusively known operator blocks or other block sparsity will also lead to reduced complexity.

\begin{figure}[htb]
    \centering
    \includegraphics[width=3.25in]{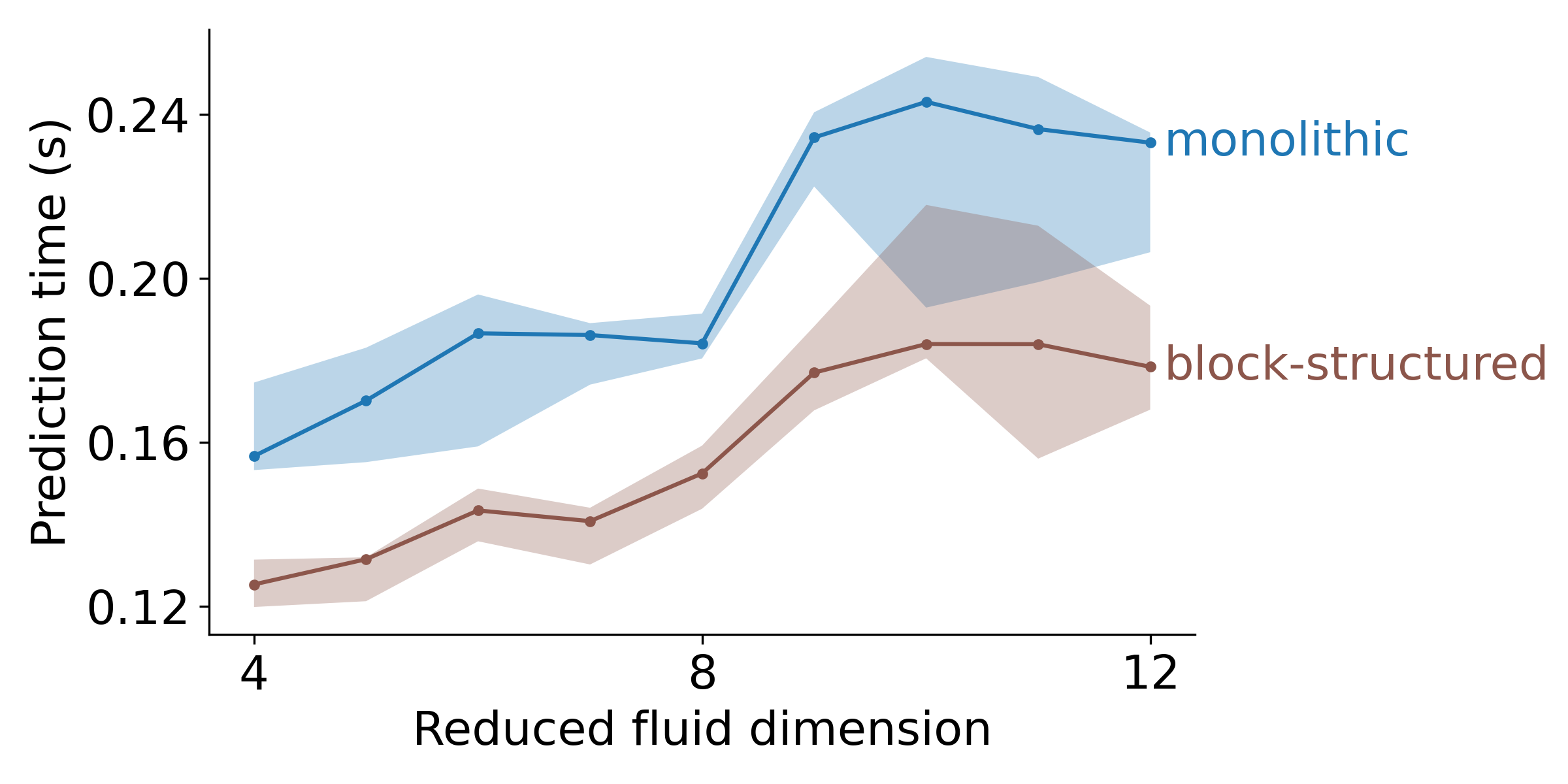}
    \caption{Prediction times averaged over flow conditions for varying reduced fluid 
 dimension with median ($50^\text{th}$ percentile) line and $25^\text{th}$--$75^\text{th}$ interpercentile shaded region.}
    \label{fig:cost_predict_k300_avg}
\end{figure}

To illustrate the practical benefits of having reduced computational complexity, Fig.~\ref{fig:cost_predict_k300_avg} illustrates the median and $25^\text{th}$ to $75^\text{th}$ interpercentile range of the online ROM prediction time across the nine flow conditions. For example, we see that at $r_\text{f} = 12$, the median prediction time decreases from 0.23 seconds using the monolithic ROM to 0.18 seconds using the block-structured ROM, approximately a 21\% decrease. This trend is consistent across all of the studied $r_\text{f}$ values (reduced fluid dimensions). Unsurprisingly, some flow conditions have more complicated dynamics and thus require longer computation times for prediction, but the relative cost between the two ROM methods remains similar across flow conditions, with the block-structured ROM performing 20\% faster than the monolithic ROM on average across flow conditions and reduced fluid dimensions for $k_\text{train}=300$.

This prediction speedup helps in two notable ways. First, the many-query setting is commonly where ROMs find their most practical engineering applications. For example, uncertainty quantification via Monte Carlo estimation typically requires many thousands of evaluations of a model to obtain a single estimate of a statistical quantity of interest. In such an application, even though the wall-clock time of the gains achieved by block-structured Operator Inference is small for a single run, the percentage gain of 20\% is high, and over an entire Monte Carlo simulation it would translate into practically useful wall-clock gains.
This is why many-query settings are able to amortize the offline cost of building a ROM by comparing to the online costs of the many predictions required after the ROM is constructed.
Second, during the regularization grid search we must infer a new Operator Inference ROM for each possible set of regularization levels and then use that new ROM to make predictions in order to identify the optimal regularization levels. That grid search can be a significant portion of the offline cost of building the ROM, so speedups can enable faster or more thorough grid searches. 

\subsubsection{Prediction outside of training set}
\label{sssec:new_ics}

We now investigate the accuracy of the monolithic and block-structured Operator Inference ROMs for perturbations (initial conditions) that are not included in the training set. We test this capability at the $M_\infty = 0.901$, $q_\infty = 70$ psf flow condition by only perturbing the initial generalized velocity $gvel_i$ for a single structural mode per test case, rather than perturbing all four modes simultaneously. We keep the initial generalized displacements $gdisp_i$ equal to zero for all four cases, but now in case 1, for example, we set $gvel_1 = 0.1$ while keeping $gvel_2 = gvel_3 = gvel_4 = 0.0$. We repeat this process for each of the four structural modes, resulting in four new test cases with a single perturbation per case.
We investigated the ROMs' accuracy for both $r_\text{f} = 8$ and $r_\text{f} = 12$, but found that both the block-structured and monolithic Operator Inference ROMs made similarly accurate predictions for $r_\text{f} = 12$, so we focus on the $r_\text{f} = 8$ ROMs here to highlight the differences.

\begin{figure}[htb]
    \centering
    \includegraphics[width=0.95\linewidth]{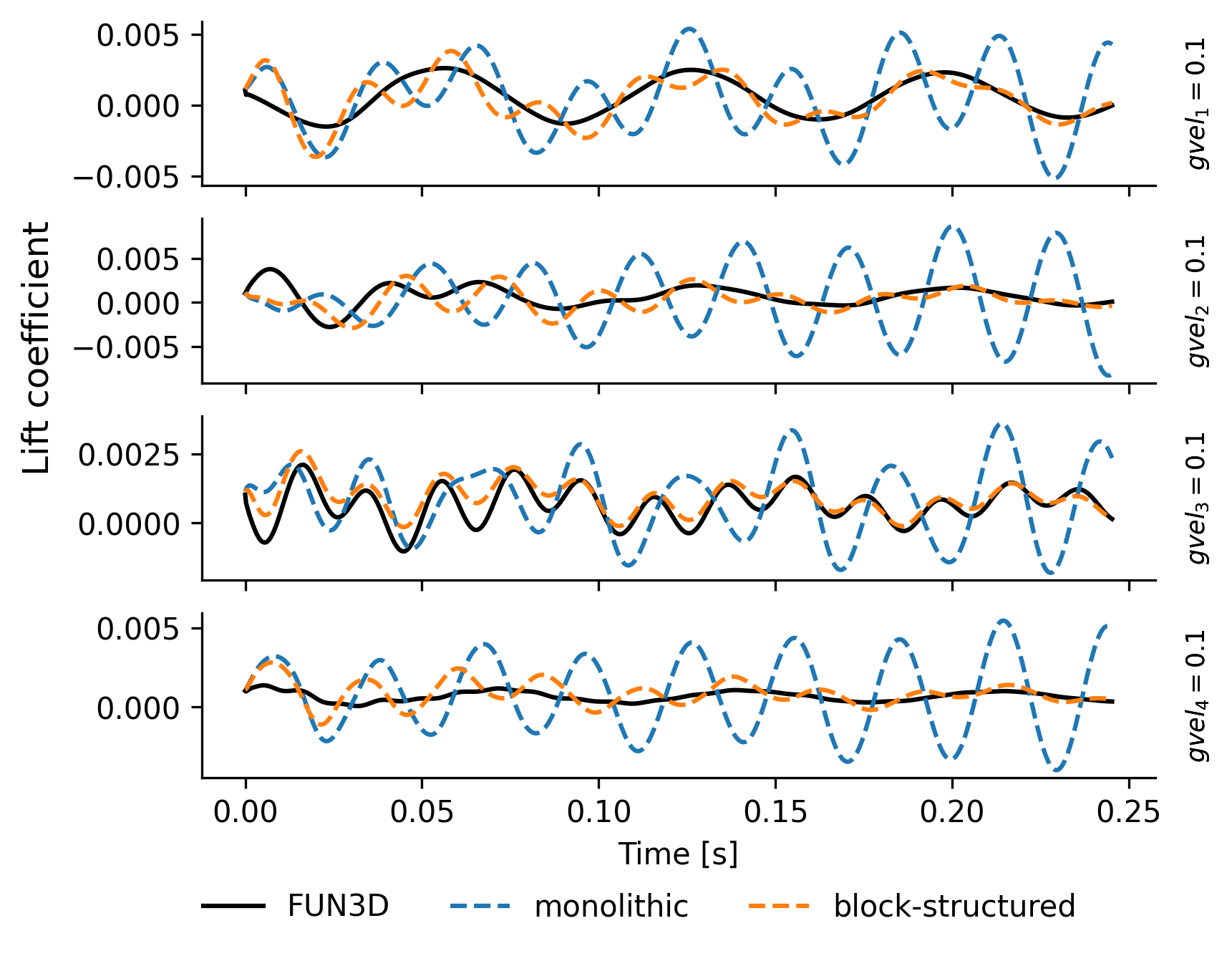}
    \caption{Lift coefficient predictions for an initial perturbation of $gvel_i = 0.1$, $i=1,\ldots,4$ with $k_\text{train} = 300$ and $r_\text{f} = 8$.}
    \label{fig:CL_multi_r8}
\end{figure}

\begin{figure}[htb]
    \centering
    \includegraphics[width=0.95\linewidth]{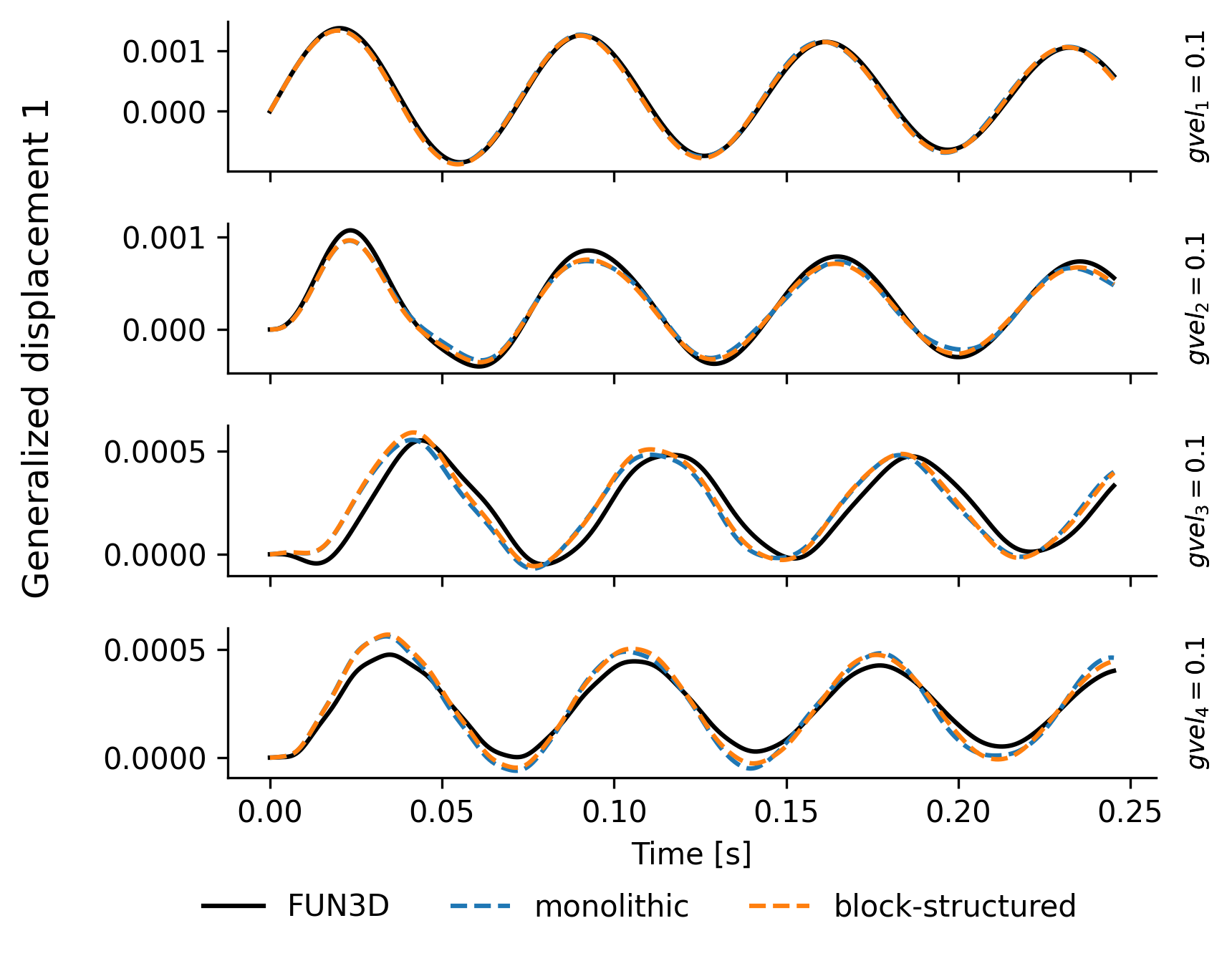}
    \caption{First generalized displacement predictions for an initial perturbation of $gvel_i = 0.1$, $i=1,\ldots,4$ with $k_\text{train} = 300$ and $r_\text{f} = 8$.}
    \label{fig:gvel1_multi_r8}
\end{figure}

\begin{figure}[htb]
    \centering
    \includegraphics[width=0.95\linewidth]{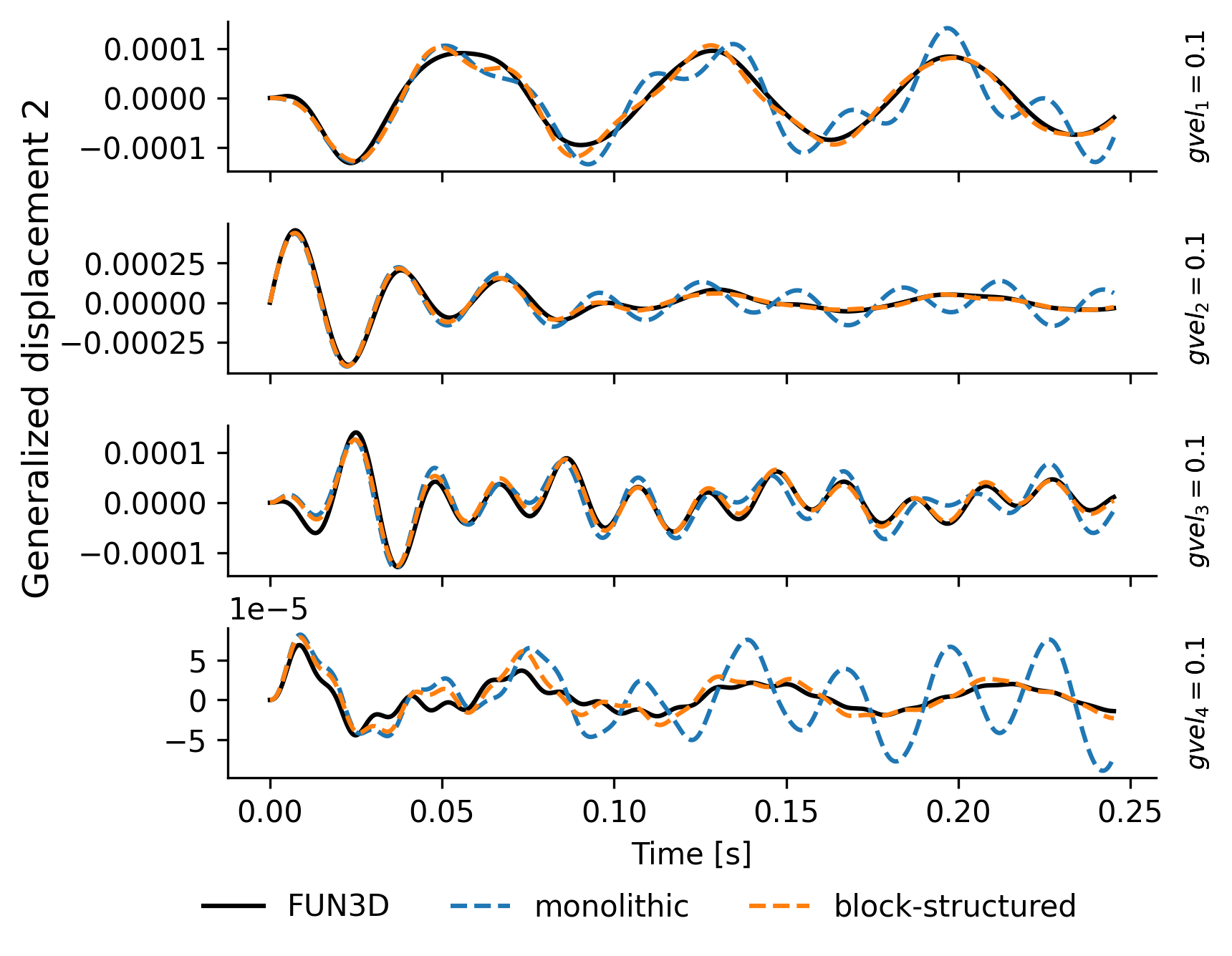}
    \caption{Second generalized displacement predictions for an initial perturbation of $gvel_i = 0.1$, $i=1,\ldots,4$ with $k_\text{train} = 300$ and $r_\text{f} = 8$.}
    \label{fig:gvel2_multi_r8}
\end{figure}

Figures~\ref{fig:CL_multi_r8}--\ref{fig:gvel2_multi_r8} show the ability of the monolithic and block-structured Operator Inference ROMs to accurately predict these new cases that are not present in the training dataset. The lift coefficient predictions in Fig.~\ref{fig:CL_multi_r8} show that the block-structured ROM is significantly better than the monolithic ROM at replicating the FUN3D-predicted lift coefficient behavior, especially as the simulations progress further in time. In contrast, the monolithic ROM struggles to generalize to the new initial conditions and we see that for all four cases the monolithic ROM incorrectly predicts that the lift coefficient oscillations will unstably increase over time.

Similar trends can be seen in the ROMs' predictions of the generalized displacements in Figs.~\ref{fig:gvel1_multi_r8}--\ref{fig:gvel2_multi_r8}. Both ROMs are able to accurately predict the response of $gdisp_1$ in Fig.~\ref{fig:gvel1_multi_r8}, but in Fig.~\ref{fig:gvel2_multi_r8} the block-structured ROM is able to correctly predict the response of $gdisp_2$ for each of the four cases while the monolithic ROM is much less accurate or even unstable. This may be due to the improved ability of the block-structured approach to precisely target the known governing equations and thus the known physics, while the monolithic ROM experiences overfitting without the additional information embedded through the coupling and block structure.

We also investigated the performance of a linear Operator Inference ROM (i.e., with no quadratic operator for either physics regime) due to the popularity of linear methods and the small magnitude of the perturbations we have imposed. We found that while the linear ROM made accurate predictions for the perturbations within the training set in Sec.~\ref{sssec:accuracy}, it struggled to make predictions for these new perturbations that were not in the training set. We have omitted these predictions from the figures for clarity, but this investigation gives us additional confidence that our quadratic, block-structured ROM is accurately predicting the nonlinear behavior of the coupled system.


% %%%%%%%%%%%%%%%%%%%%
\section{Conclusion}
\label{sec:conclusion}
% %%%%%%%%%%%%%%%%%%%%

Data-driven model reduction is often limited by a lack of physically or mathematically informed structure which can lead to overfitting and unstable dynamics. Therefore, incorporating structure into ROMs is desirable, especially in the coupled multiphysics context where problem structure plays a key role in determining the dynamical behavior of the coupled system. This paper develops the block-structured Operator Inference method to embed structure in the ROM learning problem. Taking advantage of the block structure of a multiphysics system enables separate specification of the governing equation form for each physics regime and tailored regularization choices during the least squares inference step.
The application of block-structured Operator Inference to the high-dimensional AGARD 445.6 wing aeroelastic modeling problem demonstrates the ability to learn ROMs that preserve the accuracy of monolithic Operator Inference for initial conditions in the training set, while in many cases improving upon that accuracy for initial conditions outside of the training set. The block-structured ROMs are also shown to reduce the computational expense of online prediction by an average of 20\% across flow conditions and reduced dimensions.

While the problem studied in this paper employs linear coupling, the approach admits bilinear and quadratic coupling where dictated by the underlying physical problem. The approach also enables the specification of known operator blocks intrusively. An extension of the proposed approach could incorporate parametric ROMs that further exploit the system structure exposed in the block-structured ROM.
For example, flutter prediction could be achieved by interpolation of primarily the fluid dynamics operators on a manifold, with the structural dynamics operators being parametrically dependent on the flow condition at most in the constant term, rather than interpolating all of the fluid and structural operators together.
Embedding known block structure into nonintrusive ROMs, especially for the high-dimensional multiphysics setting, provides an important step in the direction of increasing the flexibility and expressiveness of the data-driven Operator Inference method.

\clearpage
% %%%%%%%%%%%%%%%%%%%%
\section*{Appendix: FUN3D implementation of flow condition selection inspired by flutter boundary characterization}
% %%%%%%%%%%%%%%%%%%%%

Algorithm \ref{alg:flutter_params} describes the process for selecting the FUN3D input parameters, listed in Table \ref{tab:flutter_boundary_parameters}, that are needed to run an AGARD 445.6 wing simulation at a specific flow condition within the flutter characterization parameter space ($M_\infty$, $q_\infty$). These parameters were chosen to be consistent with the selected freestream mach number $M_\infty \in \mathcal{M}$ and freestream dynamic pressure $q_\infty \in \mathcal{Q}$. We require the mach number $M_\infty$ and dynamic pressure $q_\infty$ for the particular flow condition of interest and a series of constant parameters described in Tab.~\ref{tab:flutter_constants}. Table \ref{tab:fun3d_params} lists the FUN3D names for each of the required input parameters, including those computed by Alg.~\ref{alg:flutter_params}, for ease of identification and entry into the FUN3D input namelist file.

\begin{algorithm}
\caption{Selection of input parameters for a FUN3D run (originally for flutter characterization)}
\label{alg:flutter_params}
\begin{algorithmic}[1]
\Require $M_\infty$, $q_\infty$, $L$, $L_\text{nondim}$, $f_\text{char}$, $N$,
    $\gamma$, $R$, $C$, $T_\text{ref}$, $\mu_0$
\Ensure $u_\infty$, $T$, $Re_L$, $\Delta t_\text{nondim}$
\State $\rho \gets \rho(M_\infty)$
    \Comment{Fix density at original experimental flutter value in \cite{yates1987agard}}
\State $u_\infty \gets \sqrt{2 q_\infty / \rho}$
    \Comment{Compute freestream velocity}
\State $a \gets \frac{u_\infty}{M_\infty}$
    \Comment{Compute speed of sound}
\State $T \gets a^2 / (\gamma R)$
    \Comment{Compute temperature}
\State $\mu \gets \mu_0\frac{T_\text{ref}+C}{T+C}\left(\frac{T}{T_\text{ref}}\right)^{3/2}$
    \Comment{Compute viscosity via Sutherland's Law}
\State $Re \gets \frac{\rho u_\infty L}{\mu}$
    \Comment{Compute Reynolds number}
\State $Re_L \gets \frac{Re}{L_\text{nondim}}$
    \Comment{Compute Reynolds number per nondimensional unit length}
\State $\Delta t = \frac{1}{f_\text{char}}/N$
    \Comment{Compute dimensional timestep size}
\State $\Delta t_\text{nondim} \gets a \frac{L_\text{nondim}}{L} \Delta t$
    \Comment{Compute nondimensional timestep size}
\end{algorithmic}
\end{algorithm}

\begin{table}[!ht]
    \centering
    \caption{Parameter values corresponding to each flow condition in the training data generation set. Note that the time step size is nondimensional.}
    \label{tab:flutter_boundary_parameters}
    \begin{tabular}{cccccc}
        \hline \hline
        Mach number & Dynamic pressure & Freestream Velocity & Temperature & Reynolds number & Time step size \\
        $M_\infty$ [-] & $q_\infty$ [lb/$\text{ft}^2$] & $u_\infty$ [ft/s] & $T$ [\si{\degree}R] & $Re$ [-] & $\Delta t$ [-] \\
        \hline
        0.901 & 50 & 728.37 & 271.95 & 1.164e6 & 0.19826 \\
        0.901 & 70 & 861.82 & 380.72 & 1.023e6 & 0.23458 \\
        0.901 & 90 & 877.21 & 489.50 & 9.453e5 & 0.26599 \\
        0.957 & 50 & 916.65 & 381.78 & 6.856e5 & 0.23491 \\
        0.957 & 70 & 1084.60 & 534.49 & 6.186e5 & 0.27794 \\
        0.957 & 90 & 1229.82 & 687.21 & 5.813e5 & 0.31516 \\
        1.141 & 50 & 823.66 & 216.85 & 1.276e6 & 0.17704 \\
        1.141 & 70 & 974.57 & 303.59 & 1.102e6 & 0.20947 \\
        1.141 & 90 & 1105.06 & 390.33 & 1.005e6 & 0.23752 \\
        \hline \hline
    \end{tabular}
\end{table}

\begin{table}[!ht]
    \centering
    \caption{Constants used for computing flutter flow conditions for FUN3D input in Alg.~\ref{alg:flutter_params}}
    \label{tab:flutter_constants}
    \begin{tabular}{ccccc}
        \hline \hline
        \textbf{Variable} & \textbf{Symbol} & \textbf{Value} & \textbf{Units} & \textbf{Assumptions} \\
        \hline
        Characteristic length & $L$ & 1.833 & ft & AGARD wing root chord \\
        Nondimensional characteristic length & $L_\text{nondim}$ & 1.833 & - & - \\
        Frequency of interest & $f_\text{char}$ & 20.39 & Hz & Highest flutter frequency in \cite{yates1987agard} \\
        Number of timesteps per period & $N$ & 200 & - & - \\
        Specific heat ratio & $\gamma$ & 1.4 & - & Air, ideal gas \\
        Specific gas constant & $R$ & 1716.49 & ft-lbf/slug-\si{\degree}R & Air, ideal gas \\
        Sutherland's constant & $C$ & 198.6 & \si{\degree}R & Air \\
        Reference temperature & $T_\text{ref}$ & 518.69 & \si{\degree}R & Standard atmosphere, sea level \\
        Reference viscosity & $\mu_0$ & 3.737e-7 & slug/ft-s & Standard atmosphere, sea level \\
        \hline \hline
    \end{tabular}
\end{table}

\begin{table}[!ht]
    \centering
    \caption{FUN3D input parameter names}
    \label{tab:fun3d_params}
    \begin{tabular}{cc}
        \hline \hline
        \textbf{Variable} & \textbf{FUN3D parameter name} \\
        \hline
        $u_\infty$ & uinf \\
        $q_\infty$ & qinf \\
        $T$ & temperature \\
        $Re_L$ & reynolds\_number \\
        $M_\infty$ & mach\_number \\
        $\Delta t_\text{nondim}$ & time\_step\_nondim \\
        \hline \hline
    \end{tabular}
\end{table}

\section*{Funding Sources}

This work was supported in part by Lockheed Martin University Research Agreement MRA-16-005-RPP012-001, AFOSR grant FA9550-24-1-0327, and DOE grant DE-SC002317.

\section*{Acknowledgments}

We thank Kevin Jacobson and Pawel Chwalowski of the NASA Langley Research Center for their advice and insights regarding the setup of the AGARD 445.6 Wing model in FUN3D and for providing the AGARD 445.6 wing CFD grid.

\bibliography{references}

\end{document}